# A case study examining graduate student sensemaking using the epistemic game framework for Laplace's equation in upper-level electrostatics

Jaya Shivangani Kashyap and Chandralekha Singh

*Department of Physics and Astronomy, University of Pittsburgh, 3941 O'Hara St, Pittsburgh, PA, 15260*

This case study used individual interviews to investigate graduate student sense-making in upper-level electrostatics in the context of problems that can be efficiently solved for the electric potential using Laplace's equation. Although there are many technical mathematical issues involved in solving Laplace's equation, the focus of this research is not on those issues. Instead, the focus is on structural issues such as whether students recognize when solving Laplace's equation would be an effective approach to finding the potential and set up the problems correctly, and whether they can draw the electric field lines and equipotential surfaces in a given situation. Although many prior investigations have shed light on student sensemaking in the introductory physics contexts, very few investigations have focused on graduate student sensemaking while solving advanced physics problems. We present the findings of our research which was conducted through the lens of the epistemic game framework proposed by Tuminaro and Redish. We observe different nested epistemic games played by students.

# I. INTRODUCTION, FRAMEWORK AND GOAL

The research presented here is a case study that provides detailed analysis of a graduate student's sensemaking about four upper-level electrostatics problems and how they integrated physics and mathematics to solve problems that can be effectively and efficiently solved using Laplace equation. In upper-level electricity and magnetism, students typically encounter both vector calculus and differential equations in addition to other mathematical concepts they may have used previously in other physics courses. In this section, we first provide a brief review of prior research on learning electricity and magnetism, especially in advanced courses, followed by a review of research on integration of physics and mathematics, and sensemaking in physics along with synergistic frameworks.

### A. Brief Review of Prior Research on Learning electricity and magnetism

Learning electricity and Magnetism is challenging for students at all levels and requires appropriate integration of physics and mathematics [1-19]. In this context, several investigations have focused on improving understanding of electricity and magnetism among advanced physics students. For example, Singh investigated student understanding of symmetry and Gauss's law from introductory to graduate levels using a research-based instrument [20]. Bilak and Singh [21] focused on student understanding of conductors and insulators at both the introductory and graduate levels. Gire and Price [22] investigated graphical representations of vector functions in upper-division electricity and magnetism. Chasteen et al. [23] transformed their junior level course and developed an upper-level electrostatics diagnostic instrument to evaluate student learning [24], which has also been used to assess student learning of these concepts at other institutions, e.g., see Ref.[25]. Pepper et al. [26] investigated difficulties that upper-level students have with mathematics in electricity and magnetism. Wilcox and Pollock [27] investigated student difficulties with separation of variables while using Laplace's equation. Ryan et al. [28] focused on student difficulties with boundary conditions in the context of electromagnetic waves. Wilcox and Corsiglia [29] investigated student difficulties with integration across contexts in magnetostatics and classical mechanics. Bollen et al. [30] conducted investigation of student difficulties with vector calculus in electrodynamics. They investigated how students use divergence and curl in electromagnetism [31] as well as symbolic and graphical representations of vector fields [32]. Bollen et al. also developed and assessed the effectiveness of a guided-inquiry teaching-learning sequence on vector calculus in electrodynamics [33]. Maries and Singh [34] investigated physics graduate students' understanding of divergence and curl. Mason et al. [35] investigated improvement in performance in upper-division electricity and magnetism with explicit incentives to correct mistakes similar to an earlier investigation in quantum context [36].

### B. Research on Integration of Physics and Mathematics

Solving problems in physics across all levels requires integration of physical and mathematical concepts [2, 17, 28-34, 37-50]. Mathematics is not just a mere tool to solve physics problems, it is intricately intertwined with physics and is central to sense making in physics problem solving. Many times, students struggle with physics because they are developing expertise and may not activate relevant knowledge of mathematical concepts that they might have learned in their mathematics courses or may not integrate mathematical and physical concepts appropriately while doing physics. We concur with Uhden et al. [37] regarding the distinction between the technical and structural roles that mathematics plays in physics, i.e., "the technical skills are associated with pure mathematical manipulations whereas the structural skills are related to the capacity of employing mathematical knowledge for structuring physical situations". Uhden et al.'s perspective aligns with Tzanakis's viewpoint [38] that "mathematics is the language of physics, not only as a tool for expressing, handling and developing logically physical concepts, methods, and theories, but also as an indispensable, formative characteristic that shapes them, by deepening, sharpening, and extending their meaning, or even endowing them with meaning".

### C. Sensemaking in Physics and Synergistic Frameworks

Sensemaking [51-61] can be defined as the process of connecting intuitive and formal knowledge to generate new knowledge or figure out how to solve a problem. In the context of physics, when students are presented with a novel problem they have not encountered before and must solve it within a limited timeframe, they engage in sensemaking as they navigate through the problem-solving process [59, 60]. This sensemaking process can help students connect new information with their existing knowledge. As they work through the problem, students draw upon formal concepts from physics and mathematics, as well as intuitive knowledge from their everyday experiences, to bridge the gap between what they already know and what the problem is asking them to figure out. During the problem-solving process, students' epistemological beliefs about physics (e.g., whether they view it as a collection of facts and formulas and whether they perceive the instructor as the authority) can influence which knowledge resources get activated while sensemaking [62, 63]. During the sensemaking process, the activation and application of student knowledge resources to figure out how to solve the problem can lead to students combining their prior formal and intuitive knowledge to generate new knowledge. In particular, even when students' solutions aren't fully aligned with established physical laws and concepts, the act of sensemaking allows students to expand their understanding of relevant concepts beyond what they previously knew [59]. The difficulty of activating relevant mathematical as well as physical knowledge resources (even if students have them) in the appropriate physics contexts can make expert-like sense-making challenging, even in advanced physics contexts. To provide adequate support to students and help them be able to do more expert-like sense-making, it is important to understand the thinking process of students in depth when they solve problems [64-69].

Prior research shows that observing students' sensemaking can provide valuable insights into how they think while solving physics problems. It allows observation of knowledge resources that get activated in specific situations and what additional resources could be activated if students receive some guidance and scaffolding during the problem-solving process. diSessa [52, 70] introduced the concept of phenomenological primitives or p-prims, which are heuristics or mental shortcuts activated from the interaction of multiple cognitive resources, often blending intuitive elements from everyday life experiences. He and other researchers [52, 62, 70, 71] emphasized that because students' knowledge is fragmented, their sensemaking tends to show local coherence but lacks global coherence. For instance, a student may activate the 'closer is stronger' p-prim to explain why it is warmer in the summer, claiming that the Earth is closer to the Sun. In introductory physics, these researchers emphasized that students' sensemaking can be improved through support that helps them integrate their formal and intuitive knowledge resources [52, 62, 70, 71]. Tuminaro and Redish expanded the concept of p-prims into reasoning primitives, which are broader cognitive structures that operate at a more coarse-grained level of abstraction [63] and do not necessarily stem from everyday experiences. For example, the reasoning primitive 'agent causes effect' can manifest in physics as 'force as mover' or 'force as spinner' p-prims [63].

Most prior studies on sensemaking have focused on introductory physics. Our primary focus on the detailed analysis of one graduate student's sensemaking is consistent with prior studies in physics education focusing on the detailed sensemaking of a few students. For example, Dini and Hammer discussed one successful learner's epistemological framings of quantum mechanics in a case study [72]. Odden and Russ [59] illustrated a model of student sensemaking processes in introductory physics using sensemaking epistemic game. Their research [59] focused on the discussion between a pair of students discussing electricity and magnetism and suggests that "the goal of this sensemaking game is to iteratively construct an explanation in response to the perceived gap or inconsistency in knowledge". Sirnoorkar et al. [51] focused on two students and analyzed the intertwined processes of sense making and scientific modeling in the context of physics problem solving. Bing and Redish described discussions between a couple of students to analyze student epistemological framing via warrants in Ref. [53] and epistemic complexity and journeyman-expert transition in Ref. [54]. Dreyfus et al. [55] investigated mathematical sensemaking in quantum mechanics focusing on a discussion between two students.

Lenz at el. [73] developed a sophomore-level physics course that explicitly supports students in using productive sensemaking strategies while solving problems. In their case study, they described how the sensemaking of a student evolved over the course [73]. In upper-level physics, in the context of quantum mechanics, Gire and Manogue [74] investigated how two students make sense of quantum measurement, particularly the role of operators and the eigenvalue equation in sequential measurement contexts. Keebaugh et al. [50] and Hu et al. [75] illustrated students' use of several reasoning primitives in the context of quantum mechanics sensemaking more broadly and quantum computing sensemaking specifically, respectively.

Modir et al. [76] used an epistemological framing lens for characterizing student difficulties in quantum mechanics in which four frames were used: algorithmic mathematics, algorithmic physics, conceptual mathematics and conceptual physics. Gifford and Finkelstein [77] presented a framework for categorizing student sensemaking, which considers physical reasoning to understand physical or mathematical entities and mathematical reasoning to understand physical or mathematical entities as four separate modes to explain instances of student sensemaking. They identified [77] how students combine these modes through translation, chaining, and coordination. Our study presented here highlights an unexplored research area by examining graduate student sensemaking in advanced electrostatics context. We note that the grain size of our analysis of graduate student sensemaking using epistemic games presented here is larger than in Ref. [76] and Ref. [77]. At the grain size we are interested in, the four modes, frames or categories of sensemaking in Ref. [76] and Ref. [77] can often become nested or 'entangled' [78].

### D. Goal of this Study and Epistemic Game Framework

This investigation is a case study that sheds light on graduate (grad) student sensemaking during interviews conducted as a part of the development and validation of a tutorial and associated pretest and posttest on Laplace's Equation (LE) for an upper-level [24, 25, 35, 47, 74] undergraduate electricity and magnetism [23] course. We examined the sensemaking [52-54, 57, 62, 70, 79-83] of physics graduate students through the lens of the epistemic game framework, initially developed by Tuminaro and Redish [63] in the context of introductory physics problem-solving and adapted from the original framework proposed by Collins and Ferguson [84]. To the best of our knowledge, the epistemic games proposed by Tuminaro and Redish [63] have not been used to analyze sensemaking in upper-level physics courses. However, Odden and Russ [59] described a model of student sensemaking processes in introductory physics using sensemaking epistemic game while Hu et al. characterized mathematical problem solving in physics-related workplaces by interviewing those working in industries using other epistemic games [79].

This study is novel in terms of providing insight into advanced student sensemaking using the epistemic games proposed by Tuminaro and Redish [63]. Within the epistemic game framework [63], depending upon how the problem is posed, students' expertise in the domain, and their epistemological beliefs, students' cognitive control structure may activate different knowledge resources while sensemaking, and they may play different epistemic games. Tuminaro and Redish identified six types of epistemic games [63]: Mapping Mathematics to Meaning, Mapping Meaning to Mathematics, Physical Mechanism, Pictorial Analysis, Transliteration to Mathematics, and Recursive Plug and Chug. Each epistemic game has certain rules. Students can play different epistemic games following the rules as they navigate through the problem-solving process. Tuminaro and Redish [63] define an epistemic game as "a coherent activity that uses specific types of knowledge and processes associated with that knowledge to generate new knowledge or solve a problem". In physics problem-solving [80, 85-88], students engage in different epistemic games depending on the situation and the context of the problem [89]. The structural components of epistemic games include entry conditions, moves, and exit conditions. Entry conditions [63] refer to what prompts a student to begin a particular epistemic game while exit conditions mark the conclusion of the game. In problem-solving, the nature of the question can lead students to enter a specific epistemic game, while the exit conditions might involve deriving a mathematical expression, obtaining a numerical solution, offering reasoning for the problem solution, or even abandoning the problem if a student cannot figure out how to proceed. Students employ various strategies or methods, referred to as moves, to reach these exit conditions. Different types of games may require different moves based on the problem's demands.

To elaborate on our context, in electrostatics, for certain kinds of boundary value problems, Laplace's equation [$\nabla^2 V = 0$] [90] may be an effective approach to obtain the electrostatic potential $V$ in the region, where the charge density is zero. We observed the sensemaking of physics graduate students while they solved for the potential using Laplace's equation (LE) and found that this process is challenging even for graduate students. We observed that graduate students struggled to integrate mathematics with physics while doing sense-making. For example, interviews suggest that identifying problems that can be solved effectively using Laplace's equation was challenging for advanced students when they were not explicitly told which method would be effective for finding the potential in a given problem. Once students identify that solving Laplace's equation is an effective method to solve the problem posed, they can follow a standard approach to solving the partial differential equation and obtain an expression for the potential. The approach for solving problems using LE can be mathematically complex requiring multiple steps. Students can start with a conceptual analysis and represent the problem in a form, e.g., pictorial or diagrammatic representation, that can help them with the identification of the boundary conditions and the choice of an appropriate coordinate system to use before performing the separation of variables [27]. The partial differential equation in the chosen coordinate system can then be converted into ordinary differential equations by performing the separation of variables (although the focus in our investigation here is not on these later issues in the process of solving Laplace's equation). Wilcox and Pollock provided insight into how students use separation of variables to solve Laplace's equation problems in Cartesian and spherical coordinates using activation, construction, execution, reflection framework [27]. Our research, on the other hand, focuses on a case study involving grad student sensemaking. In particular, the focus here is not on issues such as the separation of variables to solve the partial differential equation for which different steps of the ACER framework used in Ref. [27] were useful.

Consistent with Uhden et al. [37], we recognize that the use of mathematics in physics can be broadly divided into structural and technical categories such that the structural usage of mathematics in physics is so entangled with the underlying physics that the two cannot be separated from each other. There are technical issues that graduate students struggled with in solving

Laplace's equation after they decided that solving Laplace's equation is an appropriate approach for finding the potential and represented the problem diagrammatically. However, in this research, our focus is primarily on their sensemaking involving structural integration of mathematics and physics. In particular, we focus here on whether graduate students recognize when solving Laplace's equation would be an effective approach to finding the electric potential and can set up the problems correctly including the boundary conditions, and whether they can draw the electric field lines and equipotential surfaces in a given situation. Due to the focus of our investigation primarily on these issues, our focus is not on symbolic forms [91]. Furthermore, we primarily spotlight the detailed sensemaking of one graduate student using the epistemic game framework but compare their sensemaking with other interviewed graduate students in some situations to highlight that they were deeply engaged in figuring things out.

## II. METHODOLOGY

### A. Interview Details

Six graduate students from a large public research university in the US were interviewed within two months. Out of these six interviewed grad students, this case study primarily centers on the sensemaking of one grad student pseudo-named G4. Three more grad students (G2, G3, and G5) are included in some cases to compare G4's problem-solving approach with other students' approaches and shed light on G4's thoughtful sensemaking. The interviewed grad students had learned these concepts related to finding the potential using various methods including Laplace's equation both at the upper-level undergraduate and graduate levels. Data were collected through semi-structured interviews conducted via Zoom, where each student was interviewed individually. The interviews were recorded on Zoom and transcribed automatically. Transcription errors were fixed by listening to the recording. The interview with each grad student comprised three distinct sessions. The first interview session included an unscaffolded pretest followed by a scaffolded pretest on LE. The pretest is given before students engage with the tutorial although they have had traditional instruction on relevant concepts at least at the undergraduate level. The unscaffolded pretest consisted of problems similar to textbook exercises without guidance, while the scaffolded pretest featured questions broken into multiple parts with incremental support. We chose to administer unscaffolded pretest first because when the problem is presented in the unscaffolded form, students are likely to struggle more in navigating the problem-solving process. The second interview session was dedicated to a tutorial designed to aid students in understanding and applying the LE method to find electric potential. The third interview session involved an unscaffolded posttest followed by a scaffolded posttest on LE, mirroring the structure of the pretest sessions but we do not discuss the posttest here.

During the interviews, students shared their screens on Zoom and used devices like an iPad/tablet to annotate a provided PDF problem set. To ensure anonymity, students turned off their videos and used pseudonyms. At the end of each session, students submitted their annotated PDFs. Each interview lasted approximately 4 hours and was divided into three sessions based on the availability of the interviewer and interviewee, with some sessions combined if necessary. Students were asked to think aloud while reasoning about the problems during the interview. The interviewer acted as an observer who did not interrupt or help except when any clarification was needed or to ask a student to try more if they were giving up easily on a problem. The collected data included Zoom recordings, audio transcripts, annotated PDFs, and narrative summaries of the sessions.

### B. Coding Scheme

The analysis employed the epistemic game framework developed by Tuminaro and Redish. This framework was used to identify and describe the epistemic games played by students during their problem-solving process. The analysis involved many iterations between the researchers including iteratively selecting and reviewing episodes of interest that demonstrated significant sensemaking and evolution in problem-solving approaches. Modifications to the ontological components of some epistemic games were made based on observations of graduate students' problem-solving processes. Our coding is structural which is a holistic approach based on the problem features [92]. After multiple iterations, both researchers discussed and agreed with the analysis of grad students' sensemaking using the epistemic game framework presented here. The first and second authors collaboratively reviewed the sensemaking episodes to ensure consistency and coherence in the analysis. Discussions focused on the local coherence in sensemaking and the relevance of the identified epistemic games. Repeated and redundant words such as 'like' and "umm" were removed from the transcript for better comprehension by the reader. Detailed analysis was performed by revisiting transcripts and video recordings of the selected excerpts for additional details. Hypotheses and interpretations were developed regarding the local coherence observed in sensemaking and the challenges associated with achieving global coherence. In this paper, we focus on a case study of G4's sensemaking as it exhibited deep engagement and notable evolution in their problem-solving strategies. They were typically more articulate than other grad students interviewed and showed a variety of interesting sensemaking patterns that can be valuable for improving physics instruction. We compare the sensemaking of G4 with other grad students in some cases to highlight G4's thoughtful approach to sensemaking and how different grad students activated different resources to solve the same problem.

### C. Questions that Prompted Sensemaking

This paper presents sensemaking in the context of four problems labeled Q1 – Q4 (for ease) in the following order: unscaffolded pretest (Q1), scaffolded pretest (Q2), and tutorial (Q3 & Q4). The reason for using this order is that we wanted to observe how the scaffolded version of the test impacts student performance compared to the unscaffolded test. In particular, before the scaffolded pretest, students were provided with the unscaffolded version of the test. Out of different questions given to students during the interviews, here we primarily focus on questions (Q1-Q4) that exemplify structural integration of physics and mathematics while sensemaking in the context of Laplace's equation. Sensemaking related to technical aspects of Laplace's equation is not discussed in this paper. Student sensemaking was observed in several sessions within a week on problems labeled sequentially as Q1- Q4. The square bracket at the beginning of each problem indicates whether the problem was part of an unscaffolded or scaffolded pretest (after traditional lecture-based instruction) or tutorial and the square bracket at the end of each problem provides a summary of the solution, which was not provided to students during interviews.

Q1. [Unscaffolded pretest] Consider the following boundary value problem: A channel consists of two parallel grounded conducting sides coinciding with the planes $y = 0$ and $y = a$. The channel is in the region $x > 0$ and is bounded at $x = 0$ with the surface set with the linear potential profile $\frac{V_0 y}{a}$ between the conducting sides, where $V_0$ is a constant. Write down the solution for the potential within the slot (channel). [This problem can be solved by solving Laplace's Equation in the channel and the given boundary conditions can be used to find the coefficients. After going through the process of solving LE, the solution for the potential in the channel is $V(x,y) = \frac{2V_0}{\pi} \sum_{n=1}^{\infty} e^{-\frac{n\pi x}{a}} \sin\left(\frac{n\pi y}{a}\right)$].

Q2. [Scaffolded pretest] For which of the following situations is solving Laplace's equation, $\nabla^2 V = 0$, an appropriate and effective method for finding the potential $V$? (Circle correct choice/choices) Explain your reasoning.
  I. A point charge $q$ is located at a distance '$d$' above a grounded conducting plane.
  II. A uniformly charged spherical volume, radius '$a$' and total charge '$Q$', enclosed within a concentric grounded conducting spherical shell, radius $b > a$, carrying charge '$-Q$'.
  III. A channel with semi-infinite conducting sides. These two sides are grounded and coincide with the planes $y = 0$ and $y = a$. They are connected in the plane $x = 0$ by a conducting sheet of width '$a$' held at potential $V = V_0$. (There is a thin insulating layer at each corner so that the conductors do not touch). [Only III]

Q3. [Tutorial] We wish to solve for the potential $V$. Circle the situations that would be best solved using Laplace's Equation, $\nabla^2 V = 0$. Explain your reasoning.
  I. A grounded conducting spherical shell of radius '$R$' in the presence of a point charge '$q$' located a distance '$d$' from the center $d > R$.
  II. A channel with semi-infinite conducting sides. These two sides are grounded and coincide with the planes $y = 0$ and $y = a$. They are connected in the plane $x = 0$ by a conducting sheet of width '$a$' held at potential $V = V_0$. (There is a thin insulating layer at each corner so that the conductors do not touch.) The channel is in the region $x > 0$.
  III. A thin disc with radius '$R$' carrying a uniform charge '$Q$' for which we want to find the potential on the axis of symmetry perpendicular to the disc. [Only II]

Q4. [Tutorial] (For Q3 II.) Make a sketch of equipotential surfaces and electric field lines within the slot (channel). You can do this using conceptual reasoning. Is the analytic solution for the potential expressed as an infinite series helpful in making the sketch? [Many aspects of student G4's sketch in the Results and Discussion section are very reasonable even though the analytic solution for the potential as an infinite series was not necessarily useful].

We note that problems Q1, Q2 (III), and Q3 (II) are mathematically well-posed problems similar to many textbook problems but are unphysical. It is not possible to create a set-up in a lab with discontinuous potential because the electric field blows up at such points. The problem is discussed for pedagogical purposes because it is possible to find the solution of the differential equations mathematically.

# III. RESULTS AND DISCUSSION

Here we present our findings regarding grad student sensemaking using the epistemic game framework. Tuminaro and Redish [63] acknowledged that the epistemic games and the possible moves they outlined for introductory physics problem-solving do not form an exhaustive list applicable in all contexts. However, they hypothesized that their framework may have broad applicability beyond the context in which they applied it. While their study focused on introductory-level students, our research centers on graduate students tackling advanced problems involving Laplace's Equation (LE) in upper-level [93, 94] electricity and magnetism. In our study, we mapped Tuminaro and Redish's proposed epistemic games in the introductory physics context onto those observed in the graduate student sensemaking about LE. However, in some instances, the ontological components of the epistemic games described by Tuminaro and Redish had to be interpreted differently for graduate student sense-making. In particular, while the general structure of the games remained consistent between introductory and advanced students, there were some key differences. For example, in many of Tuminaro and Redish's games, students primarily relied on intuitive knowledge derived from everyday experiences, often disconnected from formal academic learning. In contrast, in graduate student sensemaking regarding LE discussed here, they activated intuitive knowledge that frequently included elements learned in formal educational contexts consistent with the fact that it is difficult to directly connect abstract concepts such as how to solve differential equations for the potential with everyday experiences.

Out of the six epistemic games [63] identified by Tuminaro and Redish, we observe that in the context of grad student sense making while they solve electrostatics problems involving LE, the Pictorial Analysis game was prominent from the very beginning. Many problems are so complex that students must start with a conceptual analysis that includes a diagrammatic representation to make sense of the situation. Moreover, while sensemaking in the context of electrostatics problems involving LE, we find that students may play several nested epistemic games (e.g., the Physical Mechanism game within the move of the Pictorial Analysis game) or switch between different games while solving a single problem. For example, we find that the Physical Mechanism and Mapping Meaning to Mathematics games are sometimes played by students within the Pictorial Analysis game in a nested manner. The nesting of these games can sometimes be complex when we observe the sensemaking for a longer duration. Furthermore, sometimes the moves of the Mapping Meaning to Mathematics game involving the development/evaluation of a story about a physical situation can itself look like a standalone Physical Mechanism game.

We did not map student sensemaking in this study onto the Mapping Mathematics to Meaning game. As defined by Tuminaro and Redish [63], the difference between the Mapping Meaning to Mathematics and Mapping Mathematics to Meaning games is the structural components of the two games. The Mapping Meaning to Mathematics game begins with a conceptual story and then translated to mathematical expressions, whereas the Mapping Mathematics to Meaning game begins with a physics equation and then a conceptual story is developed around it [63]. Based on the choice of chunk/duration of the sensemaking selected, it is possible to map student sensemaking onto different types of epistemic games. We chose one way that both researchers agreed upon to map these games. However, the mapping onto various games is not unique. We note however that we expect that the nesting of the games that we observe will persist.

Tuminaro and Redish [63, 94] noted that introductory students playing the Physical Mechanism game did not use formal physics concepts (relied solely on intuitive knowledge from everyday experiences). Here our observations extend their observations to grad student sensemaking involving LE. For graduate students working on LE problems, explicitly connecting mathematical and physical concepts like integration, differentiation, and boundary conditions to everyday experiences is challenging. Instead, we find that their intuitive knowledge while solving problems involving LE often came from patterns internalized from previous formal learning in other contexts. We highlight local coherence in grad student sensemaking even if global coherence was not there, reflecting thoughtful engagement with the problem-solving process.

Table 1. Structures of the three epistemic games [63] that we observed are shown with examples from G4's sensemaking about Q1.

| Pictorial Analysis | Physical Mechanism | Mapping Meaning to Mathematics |
|---|---|---|
| Identify target concept, *e.g., identified boundary conditions and potential as target concepts in Q1 (Case A)* | Develop story about physical situation, *e.g., speculated why the surface at $x=0$ is not a conductor* | Develop story about physical situation, *e.g., evaluated that the potential due to all image charges should add up* |

| | | (equipotential surface) in Q1 (Case A) | to the total potential that satisfies boundary conditions on the upper and lower grounded conducting planes in Q1 (Case B) |
|---|---|---|---|
| Choose external representation, *e.g., drew a labeled diagram for the given charge configuration and boundary conditions in Cartesian coordinate system* | | Evaluate story, *e.g., concluded that there is a varying potential given by $V=V_0 y/a$, therefore there is some charge distribution on the surface at $x=0$ and it's not a conductor* | Translate quantities in physical story into mathematical entities, *e.g., wrote potential due to image charges by articulating distances of the image charges from the upper and lower grounded conducting planes, e.g., $V_1 = \frac{kq_1}{r_1}$* |
| Tell a conceptual story based on spatial relations among the objects, *e.g., formulated a story about what the potential should be at different locations based on the given configuration* | | | Relate mathematical entities in accordance with physical story *e.g., ensured that boundary conditions are satisfied with total potential on each grounded conducting plane, e.g., $V_{total} = V_1 + V_2 + ...$ where $V_1 = \frac{kq_1}{r_1}$, $V_2 = \frac{kq_2}{r_2}$ etc. with appropriate distances* |
| Fill in the "slots" in the representation, *e.g., obtained a labeled diagram with boundary conditions specified* | | | Manipulate symbols *e.g., $V_{total} = \frac{kq_1}{r_1} + \frac{kq_2}{r_2} + \cdots$* |
| | | | Evaluate story *e.g., concluded total potential satisfies the boundary condition on the upper and lower conducting planes and the channel consists of equipotential surfaces parallel to the conducting planes consistent with $V=V_0 y/a$* |

### A. Case A. Conflict about values of potential on two parallel grounded conducting planes

We begin with the discussion of student G4's sensemaking while solving Q1. Here we focus on G4's sensemaking early in their problem-solving process while doing a qualitative analysis of Q1. We also compare G4's sensemaking with another grad student G3 to observe the similarities and differences in the activation of various knowledge resources while problem-solving. We note that G4 spent significantly more time in sensemaking on Q1 compared to G3.

The given problem Q1 was unscaffolded and the goal of the problem was to obtain an expression for the potential within the channel. G4 read Q1, recognized the need for a diagrammatic representation of the given configuration to clarify the situation and entered the pictorial analysis game (see Table 1 first column). Their identified target concepts in the game were boundary conditions and the potential. They chose to draw a diagram using Cartesian coordinates as an external representation to visualize the given configuration. They started drawing the diagram shown in its final form in Fig. 1. They said, '...*So, we're going to draw that one* [configuration] *right out...so we have this...*[started drawing the $x$ and $y$ axis] *$y$ is between '0' and 'a'* [marked the dots on the $y$ axis]...*The channel is in the region $x > 0$* [drew two parallel lines showing $y = 0$ and $y = a$ planes]...*and is bounded at $x = 0$...So, then there's another boundary right here* [drew a vertical surface at $x = 0$ plane] *with the surface set with the linear potential...*'. G4 then wrote down the expression for the potential profile on the vertical surface saying, '*at $x =$*

$0$, *the potential profile is* [wrote down] $\frac{V_0 y}{a} = V(x = 0, y)$.' G4 shaded the region within the channel to emphasize the region where the potential is to be found, saying, '*So I'm assuming we're only in two dimensions…Whenever they say potential within the slot* [channel]*…I assume they mean anywhere in here* [shaded the region between the bounded surfaces]'.

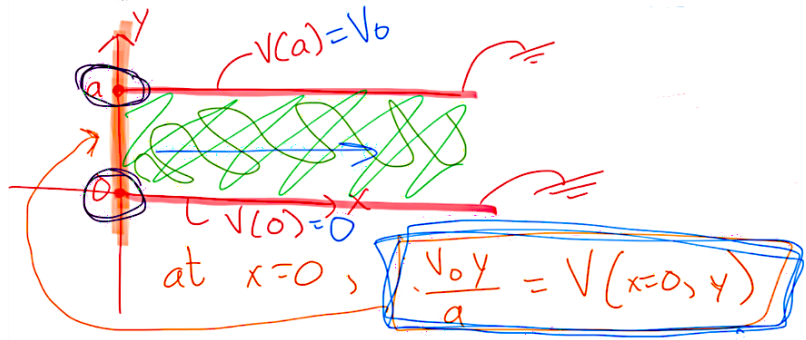

Fig. 1. First diagram drawn by G4 for Q1 while analyzing the problem

After completing the basic components of the diagram [Fig. 1], they contemplated the potential on the grounded conducting planes and thought that this would help them in finding the potential within the channel. G4 noted that the potential on the two grounded conducting planes, $y = 0$ and $y = a$, should be related to the potential $\frac{V_0 y}{a} = V(x = 0, y)$ (although both conducting planes were grounded and must have the same potential). This perceived need to understand the values of the potential at different grounded conducting surfaces and resolve any contradiction, motivated G4 to enter the Mapping Meaning to Mathematics game within the Pictorial Analysis game (as they were not completely done drawing the final diagram in Fig. 1). As the first move of the Mapping Meaning to Mathematics game, they started analyzing the potential at the two grounded parallel conducting planes (which helped them in developing a story about the physical situation in the given configuration).

G4 drew grounding symbols on the parallel conducting planes consistent with the fact that those planes are grounded but incorrectly concluded that the two grounded conducting planes can have different constant potential values based on the potential profile of the vertical surface at $x = 0$. They said, '*…A channel consists of two parallel grounded conducting sides* [read the question again] *…Okay, so then, that tells me that right here this is grounded,* [drew the grounded sign at the upper plane and then the lower plane]*…and that's grounded. Which means that the potential is going to be some value at the top. I'll call this $V(a)$* [marked it on the upper plane] *and some potential at the bottom, this is $V(0)$* [marked it on the lower plane]*…So this value* [referring to $\frac{V_0 y}{a} = V(x = 0, y)$] *is only at $x = 0$. So that's only for this line here* [referring to surface at $x = 0$]'. Since $y = 0$ and $y = a$ planes and the surface at $x = 0$ are in contact, G4 wondered whether their potential values should match. Their confusion is reasonable because, at $x = 0$, the infinitesimal shift in the position completely changes the potential at the upper grounded conducting surface. G4 said, '*…Okay, so I guess this is* [referring to $\frac{V_0 y}{a} = V(x = 0, y)$] *…boundary condition number one. Alright, and I guess the farther away from this* [vertical surface at $x = 0$] *we go, it should only be determined by the potential at $y = 0$ and the potential* [at] *$y = a$…So normally, we would set up some sort of three-way boundary condition…The potential at $x = 0$ and $y$ is $V_0 \left(\frac{y}{a}\right)$. So at $y = 0$, potential is equal to $0$ and at $y = a$, potential is equal to $V_0$…*'. G4 wrote down the boundary conditions as $V(x = 0, y) = V_0 \left(\frac{y}{a}\right) \rightarrow at\ y = 0, V = 0, at\ y = a, V = V_0$.

The mathematical expressions they obtained for the potential at the junction of the $x = 0$ plane and the two grounded conducting planes *at $y = 0$, i.e., $V = 0$, and at $y = a$, i.e., $V = V_0$* did not completely convince them that they were correct. They contemplated whether $V_0$ should be the potential just at $x = 0, y = a$ or over the entire upper grounded conducting plane. They said, '*…Okay, so I can write those values…in here at $y = a$, $V_0$* [labeled the upper grounded plane in Fig. 1 as $V(a) = V_0$]*, and then over here we're at zero* [labeled the lower grounded conducting plane in Fig. 1 as $V(0) = 0$]*. Okay? So, it doesn't tell us if this* [surface at $x = 0$]*, whatever it's bounded by at $x = 0$ is a conductor…Okay…we know for a fact…at $x = 0, y = 0$, potential is $0$,* [and] *at $x = 0, y = a$, potential is $V_0$…Is this* [true that for] *$x > 0$, $y = a$,* [potential is] *$V = V_0$? Same with $x > 0$, $y = 0$, is* [potential] *$V = 0$?…We know what it* [potential] *is at these two points* [at $x = 0, y = 0$ and $x = 0, y = a$]*…I*

*guess we're allowed to say that that whole entire plate* [referring to each grounded conducting plane] *is at the same potential because…it's grounded*'.

G4 now proceeded to do sensemaking about whether the surface at $x = 0$ is a conductor and decided to enter the Physical Mechanism game (Table 1 middle column). We note that the Physical Mechanism game in this case can be considered a part of the last move of the Mapping Meaning to Mathematics game, which consists of evaluation of the story. The story here is related to identifying the nature (conductor/non-conductor) of the surface at $x = 0$. They said, '*…It says that there's a surface with the linear potential between the two conducting sides* [parallel grounded planes]. *But to my knowledge, it doesn't tell us what that surface is at $x = 0$…because…it can't be a conductor. Because if it was a conductor, then it would all have the same potential…I guess if there's a certain potential at $x = 0$, $y = a$ it would be the same potential for that whole thing* [$y = a$ grounded conducting plane]. *So, in that case, I don't know why this* [referring to $\frac{V_0 y}{a} = V(x = 0, y)$] *isn't just the potential everywhere within it* [channel]*…I guess there's some charge distribution along that y-axis…There has to be, because, if* [it is] *not a conductor, there is a change in potential for one point to another…I know…there's some charge distribution…along the y-axis and that whole potential is the same everywhere on $y = a$ and $y = 0$ lines* [referring to the grounded conducting planes].

We observe that G4 eventually convinced themselves that the potential should be the same over the entire upper grounded conducting plane at $y = a$ although it is a different value from $V=0$ on the lower grounded conducting plane (which is not correct). While figuring out the potential on the boundaries, G4 also concluded that the vertical surface at $x = 0$ could not be a conductor and there should be some charge distribution on that $x = 0$ surface. This was a productive line of thought since the charge distribution on this surface was not explicitly mentioned (or asked for) in the problem statement of Q1. With a physical understanding of the given configuration in Q1 to their satisfaction, they exited the Physical Mechanism game. At this point they had also figured out that the values of the potential on $y = 0$ and $y = a$ grounded conducting planes should be $V = 0$ and $V = V_0$, respectively. Therefore, they exited the Mapping Meaning to Mathematics game. At this point in their sensemaking, they also exited the Pictorial Analysis game, being satisfied with their detailed and labeled diagram of the given configuration. The summary of the different nested epistemic games played by G4 in Case A is shown in Fig. 2. At this point, G4 focused on sensemaking to find the solution within the channel in Q1, which is discussed in Case B.

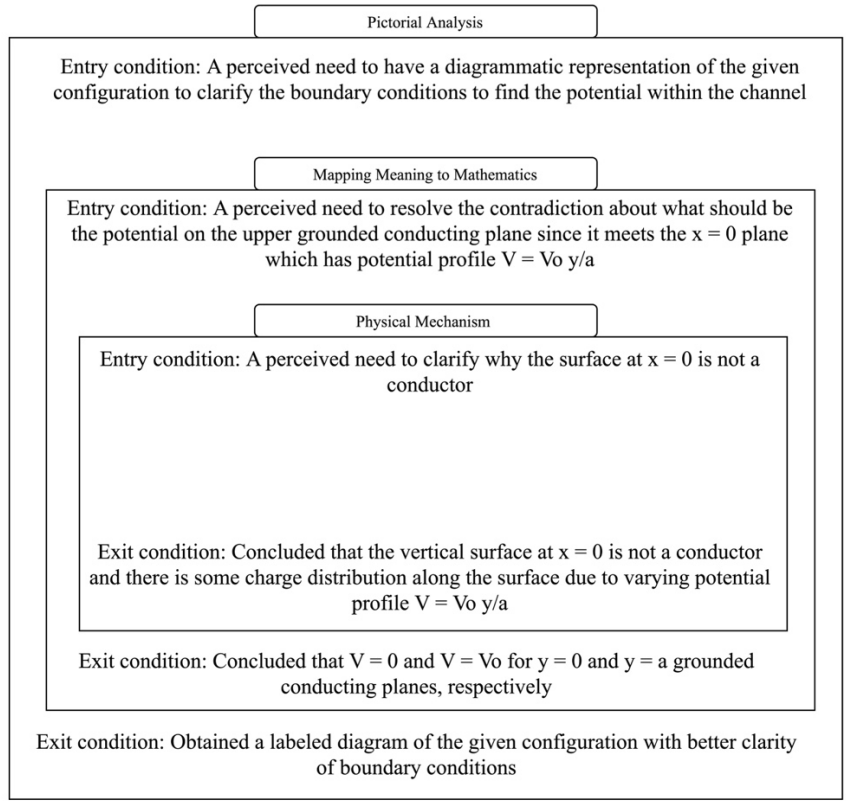

Fig. 2. Summary of different nested epistemic games played by G4 in Case A

For comparison, we now discuss the sensemaking of another grad student G3 about Q1, who also got confused about the potential at the two grounded conducting planes similar to G4. As G3 read Q1, they perceived the need to draw a diagram of the given configuration and entered the Pictorial Analysis game. Their identified target concepts in the game were boundary conditions and the potential. G3 chose to draw a diagram using Cartesian coordinates for the given configuration. They started drawing the diagram which is shown in the final form in Fig. 3. G3 said, '*I'm gonna try to draw a picture, because there's a lot of positions and stuff* [information] *specified. And I can't really visualize that without the picture*'. They started drawing a diagram with $x$ and $y$ axes, '*…This is only the xy plane, so I'm gonna go ahead and draw this* [drew the $xy$ plane]. *Okay, we have two parallel grounded conducting sides at…$y = 0$ and $y = a$* [drew planes]*…The channel is then in the positive x region. So, it's over here bounded at $x = 0$…with the surface set with a linear potential profile between the conducting sides.*'

Once G3 drew the diagram in Fig. 3 with the relevant axes and grounded conducting planes, they wanted to find the potential within the channel given that the potential profile on the vertical $x = 0$ surface is $V = \frac{V_o y}{a}$. The need to write the potential between the parallel conducting planes at $y = 0$ and $y = a$ made G3 enter the Mapping Meaning to Mathematics game within the Pictorial Analysis game. As the first move of the game, they focused on the given potential profile $V = \frac{V_o y}{a}$ on $x = 0$ surface. They said, '*So that's* [the potential $\frac{V_o y}{a}$] *in between* [the two grounded conducting planes]. *Here is the $V = \frac{V_o y}{a}$ I believe* [they incorrectly thought $V = \frac{V_o y}{a}$ represents the potential profile everywhere in the channel shown in Fig. 3 instead of only at $x = 0$ as given], *I think that's what the problem is saying*'.

This proposed solution within the channel is simplistic because G3 claimed that the potential at $x = 0$ (given in the problem statement) is the potential everywhere in the channel that the problem asks for. They spent a very short time verifying their proposed solution, an attempt that did not involve deep sensemaking. Instead, they either re-read the problem or repeated the same thing they had already said earlier. They said, '*Write down the solution for the potential within the slot* [channel]*…So, within the slot* [channel]*, well, okay, so I'm given this* [$V = \frac{V_o y}{a}$] *value as a function of distance…So this* [potential within the channel] *is not necessarily constant because y changes…But I think this* [$\frac{V_0 y}{a}$] *tells me* [started writing the potential for bottom grounded conducting plane] *at y = 0…potential is zero and then that's the other plane* [$y = a$ grounded conducting plane], *it's just constant* [but not zero, i.e., $V_0$]…*So I noticed…this* [$\frac{V_0 y}{a}$] *is not a function of x position, it's a function of y position. So…the solutions on these two surfaces…would be…*[wrote $V = 0$ and $V = V_0$ for $y = 0$ and $y = a$, respectively, as shown in Fig. 3]. *So, I think that's all…I'm really gonna stick with these two values* [$V = 0$ and $V = V_0$ on grounded conducting plane] [and $V = \frac{V_0 y}{a}$ within the channel] *as my solutions*'.

Once they proposed the solution within the channel as $V = \frac{V_0 y}{a}$ and boundary conditions at $y = 0$ and $y = a$ as $V = 0$ and $V = V_0$, respectively, they exited the Mapping Meaning to mathematics game without spending much time in the move of the game involving evaluation of the story. Since they also were satisfied with their labeled diagram, they also exited the Pictorial Analysis game.

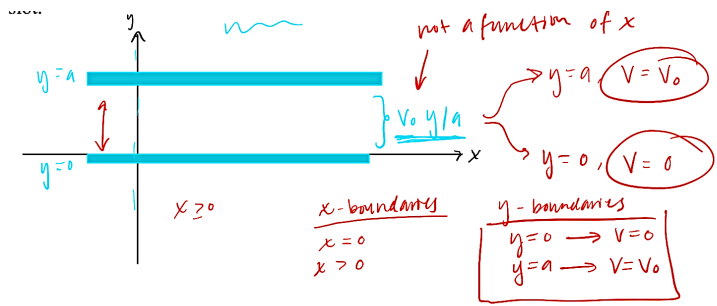

Fig. 3. Diagram drawn by G3 for Q1

Comparing the two grad students, although both G3 and G4 arrived at the same solution, G3's sensemaking is not as deep as G4's sensemaking. For example, due to the potential profile on the vertical $x = 0$ surface being $V = \frac{V_0 y}{a}$ and the vertical $x = 0$ surface meeting the horizontal upper and lower grounded conducting planes, G4 perceived a conflict, e.g., whether the potential on the two grounded conductors should be different consistent with $V = \frac{V_0 y}{a}$ or be the same. G4's sensemaking was thoughtful as they tried to resolve the conflict by reasoning, e.g., about what the potential on the upper grounded conducting plane should be in Fig. 1, while analyzing the problem and setting up the boundary conditions. On the other hand, without worrying about whether the potential on the lower and upper grounded conducting planes could be different, G3 concluded that the potential on those grounded conducting planes are different ($V = 0$ and $V = V_0$ for $y = 0$ and $y = a$, respectively, consistent with $V = \frac{V_0 y}{a}$ at the vertical $x = 0$ plane). G3 also noted that the potential $V = \frac{V_0 y}{a}$ at the vertical $x = 0$ plane does not depend on $x$ and quickly concluded that the potential within the entire channel which the problem asked for is the same as $V = \frac{V_0 y}{a}$ (which is somewhat simplistic). At this point, G3 assumed they were done solving the problem without worrying about how the problem solution could be trivial. For contrast, G4's sensemaking of the linear potential profile at $x = 0$ being due to some charge distribution on that surface was an excellent example of meaningful inference that was not explicitly asked for in the problem. This inference related to the charge distribution on the $x = 0$ surface was at least partially motivated by the fact that they were trying to resolve contradictions they had encountered while doing sensemaking regarding the boundary conditions. They wanted to resolve various conflicting ideas they activated while setting up the problem, e.g., whether the potential on the upper grounded conducting plane should be (a) zero because it was grounded, (b) whether it should be $V_0$ only where the upper conducting plane touched the $x = 0$ surface and zero elsewhere or (c) whether it should be a non-zero value $V_0$ on the entire upper conducting plane since the conductor is an equipotential surface in electrostatics. Although they finally incorrectly convinced themselves that the potential $V_0$ on the entire upper grounded conducting plane should be the same value $V_0$ and different from the lower grounded conducting plane at zero potential, their sensemaking to resolve these conflicting ideas is praiseworthy.

### B. Case B. Blending Laplace's equation with the method of images

Case B is a continuation of the sensemaking of G4 for Q1 discussed in Case A. G4 contemplated the boundary conditions for the upper conducting plane in Q1 and introduced the idea that there must be charge distributed on the vertical surface at $x = 0$, where the potential is linear in the $y$ coordinate, i.e., $V = \frac{V_0 y}{a}$. At this point, G4 felt that they had completed their initial conceptual analysis and set up the problem diagrammatically so that they could now focus their attention on planning the solution for Q1. G4 looked at their diagram and thought that applying the method of images would be appropriate to solve this problem (the configuration in problem Q1 is not suitable for the method of images). G4 did not explicitly mention why they activated knowledge resources pertaining to method of images for Q1 but we hypothesize that their focus on resolving the conflicts regarding the two grounded conducting planes could be a possible reason (since grounded conducting planes are often important considerations in the method of images).

The activation of resources that the method of images is suitable for Q1 prompted the need to have image charges in the given configuration and made G4 to enter the Pictorial Analysis game again. Their identified target concepts in the game were image charges and the potential within the channel. They chose to draw a diagram consisting of image charges, using Cartesian coordinates as an external representation. It was a new diagram as shown in Fig. 4 in which there was also more space to place the image charges they thought were needed to solve the problem (compare Fig. 1 and Fig. 4). Fig. 4 was drawn for a better interpretation and clarity on how to proceed with the problem-solving process and for the placement of the image charges. They said, '*...I'm gonna break it* [diagram in Fig. 4] *up into 3 pieces…There's that line* [drew the bottom line as shown in Fig. 4, representing the $y = 0$ (semi-infinite) plane], *there's this line* [drew the top line as shown in Fig. 4, representing the $y = a$ plane] *and then there's this line* [drew a vertical line representing the surface at $x = 0$]. *On that line* [vertical line representing the surface $x = 0$], *potential is equal to* $V = \frac{V_0 y}{a}$, *on the red line* [representing the $y = 0$ plane], $V = 0$ *and then on the blue line* [representing the $y = a$ plane], $V = V_0$. G4 continued using the same boundary conditions, which they marked in Fig. 1.

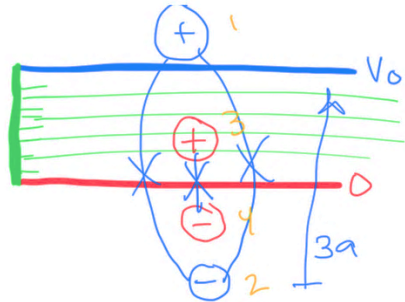

Fig. 4. Second diagram with image charges drawn by G4 for Q1 while planning the solution

Now they wanted to figure out how many image charges should be placed and where they should be placed such that all boundary conditions are satisfied (although this is not a correct method). To justify their configuration of image charges physically, they started playing the Physical Mechanism game nested within the moves of the Pictorial Analysis game (as they were still working on the diagram). The story in the game was related to the placement of image charges in the given configuration, satisfying boundary conditions. They said, '*…I don't know how you could do that* [have potential $V = 0$, on the $y = 0$ plane and $V = V_0$ on $y = a$ plane in the presence of image charges]*…Looking at the red line, we could have a positive charge here* [placed a positive image charge above the $y = 0$ plane as shown in Fig. 4] *and negative charge here* [placed a negative image charge right below the $y = 0$ plane], *and it would cancel itself out if we were doing like a method of images thing. And then we could probably even do another positive charge over here* [placed a positive image charge above the $y = a$ plane], *and maybe another negative charge over here* [placed a negative image charge below previously placed negative image charge]. *It* [overall charge configuration] *would still cancel out what's* [potential] *on the red line* [$y = 0$ plane] *because these two* [first two image charges placed by them] *would cancel out, and then these two* [last two image charges placed by them as shown in Fig. 4] *would…cancel each other out. So a red line* [grounded conducting plane at $y = 0$] *would still be at potential equal to 0…But then, if we were to look at this blue line* [$y = a$ grounded conducting plane], *there's two positive charges* [referring to two positive image charges as shown in Fig. 4] *near it, and two negative charges really far away* [referring to two negative image charges as shown in Fig. 4]. They placed the image charges such that all boundary conditions are satisfied to their satisfaction.

G4 said, '...*So if we were to think about...a point charge, potential is equal* [wrote down] $V = \frac{kq}{r}$. *So the blue line* [$y = a$ plane] *would have some positive charge near it and the red line* [$y = 0$ plane] [potential] *would still be canceled out* [$V = 0$ on lower grounded conducting plane], *as long as these distances were the same* [referring to the distances between positive and negative charges they placed from the lower grounded conducting plane].'

We note that G4 used $y = 0$ as the reference line to measure the distances to the charges they placed as shown in Fig. 4. They said, '...*from the red line* [$y = 0$], *I should say...the two red charges* [first pair of positive and negative image charges placed by G4] *are the same distance away from each other. The two blue charges* [last pair of positive and negative image charges placed by G4] *are the same distance away from each other, in respect to the red line*'. They started writing the position of image charges with respect to the $y = 0$ plane saying, '...*blue charges* [last pair of positive and negative image charges respectively] *are at $y > a$ and $y < -a$, but same magnitude...The charges would be whatever they need to* [be to] *make sure that it's $V_0$ at the blue line* [$y = a$ plane] *and then zero down here* [$y = 0$ plane]'.

G4 said that surfaces parallel to $y = 0$ (as drawn in Fig. 4) are all equipotential surfaces. At this point, G4 thought about equipotential surfaces within the channel but did not put too much emphasis on this idea. They said, '...*And of course, linearly* [going horizontally on one of the green lines representing an equipotential surface], *wherever we end up along this line* [pointing to the lines parallel to $y = 0$ plane as shown in Fig. 4], *it's gonna be the same* [potential], *no matter where we are in here.*'

As G4's focus was on placing the image charges such that the boundary conditions were satisfied, they did not dig deep into the idea of equipotential surfaces within the channel at this time. They started focusing on finding the positions of image charges measuring them from the $y = 0$ plane. G4 labeled all these image charges as shown in Fig. 4 saying, '...*So I'm going to call these charges one* [positive image charge at the top], *two* [negative image charge at the bottom], *three* [positive image charge between the $y = 0$ and $y = a$ plane], *and four* [negative image charge right below the $y = 0$ plane]'. They wrote down the positions of these image charges saying according to the position labels, '...*one* [image charge] *is at $y = 2a$, two* [is at] *$y = -2a$...three* [image charge] *is at $y = \frac{1}{2}a$, and then four* [is at] *$y = -\frac{1}{2}a$, because they just need to be the same distance away from that red line* [$y = 0$ plane]. *That should make sure that the red axis* [$y = 0$ plane] *is always at some potential $V = 0$, and it should make sure that blue axis* [$y = a$ plane] *is always at some other potential* [$V_0$]'.

Once G4 obtained four image charges which seemed to be physically satisfying the boundary conditions according to them, they exited the Physical Mechanism game.

Although G4 was convinced that the boundary conditions were being satisfied with the four image charges physically, they wanted to ensure that it is true mathematically, particularly for the $y = a$ plane. This need to verify boundary conditions at $y = a$ plane mathematically made them enter the Mapping Meaning to Mathematics game. As a part of the development of the story, they started analyzing the contributions to the potential from all the image charges at the $y = a$ plane. They decided to write an expression for the potential on the $y = a$ plane. G4 said, '*Let's...write out the potential along that blue axis* [$y = a$ plane as shown in Fig. 4]. *At $y = a$, the potential is going to be the potential from the first* [image charge], *plus the potential from the second charge, plus the potential of third plus the potential of the fourth* [wrote that at $y = a \rightarrow V = V_1 + V_2 + V_3 + V_4$]'. They said, '*Potential from charge one, we're at a* [writing down potential at $y = a$], *that other charge is at a distance of 'a' away* [from $y = a$ plane] *because it's '2a'* [from $y = 0$ plane], *so it's $\frac{kq}{a}$... Then potential* [due to charge] *two is negative, so we have $kq$ over the distance...for charge two, it's -2a all the way up to here* [G4 pointed to the distance of the lower grounded conducting plane $y = 0$ from charge two], *so that's '3a'* [distance from $y = a$ plane to charge two]...*Do we care about...negative or positive direction? I don't think we do. No, because the charge takes care of that...Okay, and then for charge three...that distance is $\frac{a}{2}$, and charge three is positive, that's gonna be $+\frac{kq}{a/2}$. And then the last one, charge four is negative...is one and a half 'a' away* [from $y = a$ plane], *so it's three halves 'a'*...[while saying this, they wrote down the potential at '$a$' as $V(y = a) = \frac{kq}{a} - \frac{kq}{3a} + \frac{2kq}{a} - \frac{2kq}{3a} = V_0$]...*because we know that this whole thing is equal to $V_0$...I'm just checking boundary conditions and making sure it looks good. So we have* [they wrote] $kq\left[\frac{3}{3a} - \frac{1}{3a} + \frac{6}{3a} - \frac{2}{3a}\right] = V_0$.'

G4 started writing the equation as an attempt to verify that the boundary condition is being satisfied at $y = a$ plane due to the image charges and they inadvertently ended up generating an expression for $V_0$. Then they wrote $V(y = a) = \frac{6kq}{3a} = V_0 = \frac{2kq}{a}$.

We note that G4 stated that the upper grounded conducting plane is an equipotential surface with potential $V_0$. However, they did not check whether the expression for the potential at all points on the upper grounded conducting plane would work out to be the same based upon their positioning of their image charges. In particular, their calculation was only for a point on the upper conducting plane along the straight line joining the image charges. Thus, knowledge resources were not activated pertaining to the fact that the distances to the image charges in Fig. 4 would be different for different points on the upper grounded conducting plane (so the potential will not necessarily be the same at all points on that plane due to the image charges).

At this point, G4 realized that they had still not found the potential within the channel that the problem asked for. This thought made them go back to the idea of equipotential surfaces within the channel, which they contemplated briefly earlier. The need to understand how the potential varies within the channel physically made them enter the Physical Mechanism game within the Mapping Meaning to Mathematics game. They developed a story by comparing the given configuration to a parallel plate capacitor. They had already drawn the equipotential lines inside the channel. At this point, they thought that the potential is linear and said, '...*it's linear...everywhere inside of there* [channel], *it's just going to be linear because it's almost like we have two parallel plate capacitors*'. As G4 was playing other games within the Pictorial Analysis game, they redrew two parallel lines (not shown here) saying, '...*Let me redraw this. So up at the top, we have this* [drew a line representing plane $y = a$]...*and then down at the bottom, we have this* [drew a line representing plane $y = 0$] *and then we're at* $V_0 = \frac{2kq}{a}$ [value of the potential at the $y = a$ plane they had found although it is not correct]. *Then here we're at* $V = 0$ [value of potential at $y = 0$ plane]. *The potential* [is] *sliding between them* [value of potential within the channel], *potential at any 'y' is going to be* $V(y) = \frac{V_0 y}{a}$. As G4 had already obtained an expression for $V_0$, using image charges, they thought it would be useful to use that value in the final expression for the potential. They concluded that the potential within the channel can be written as $V = \frac{V_0 y}{a}$ with $V_0 = \frac{2kq}{a}$. With this inference, G4 exited the Physical Mechanism game. They said, '*But now I can just write it* [potential within the channel] *more simply,* $V(x > 0, y) = \frac{2kq}{a} \cdot \frac{y}{a}$, $V(x > 0, y) = \frac{2kqy}{a^2}$. *And this is my final answer*'. Once G4 obtained the final expression for the potential, they exited the Mapping Meaning to Mathematics game. While doing this calculation, G4 appeared to be confident about their answer and thought that they had solved Q1 correctly, because their expression for the potential looked justifiable to them even though it was not correct. Since G4 was done with all the diagrams, they also exited the Pictorial Analysis game. Fig. 5 shows a summary of the different nested epistemic games played by G4 in Case B.

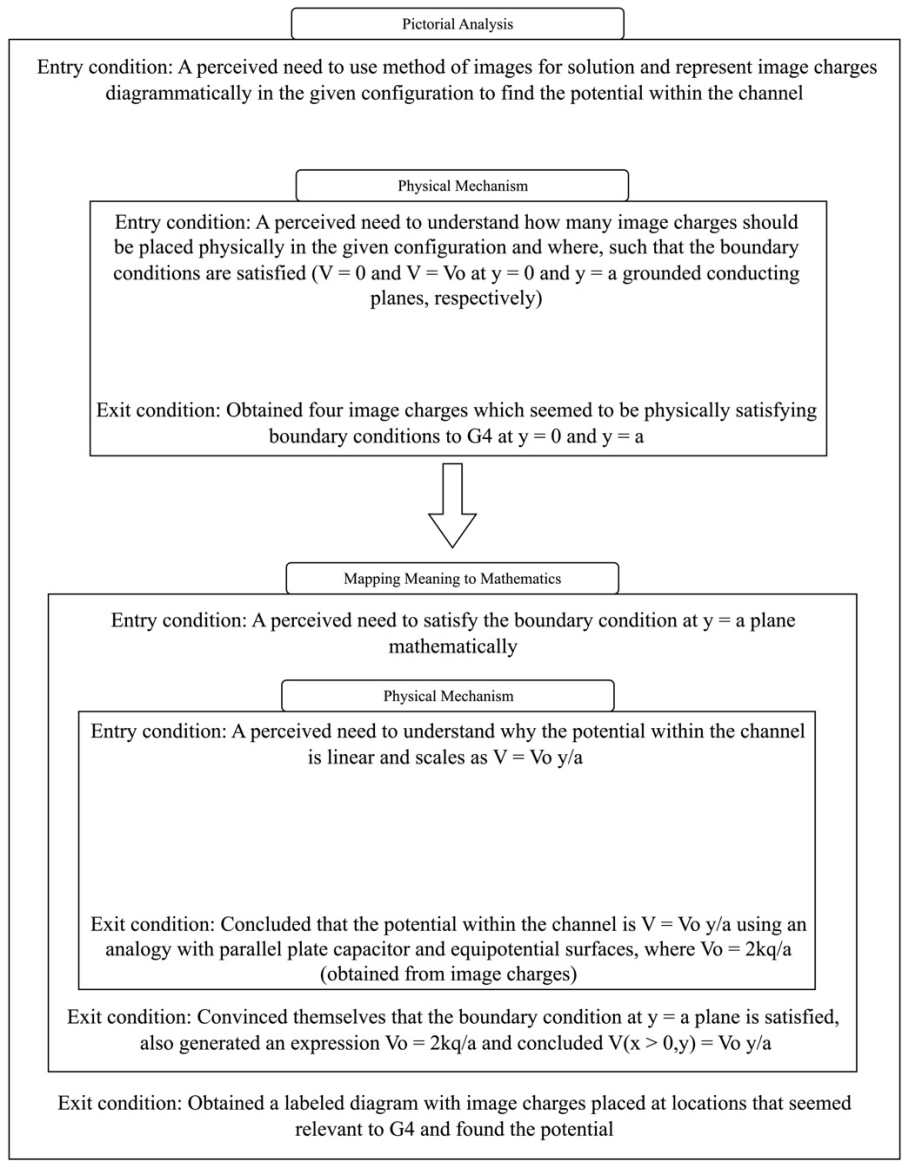

Fig. 5. Summary of different nested epistemic games played by G4 in Case B

In this example, we observe that while contemplating how to solve Q1 after convincing themselves that the potential on the upper grounded conducting plane $V_0$ is different from zero, G4 proceeded to use the method of images to solve the problem by placing image charges (when the method of images is not applicable here). The image charges they placed in Fig. 4 did not correspond to point charges that are present in the problem as should be the case. We note that in this problem (in which the problem statement does not explicitly say which method should be used), G4 blended different methods while sensemaking. This blending of different concepts and activation of a combination of appropriate and inappropriate knowledge resources in a particular context is likely to happen when students learn different methods and concepts in a short duration of time. The different approaches, e.g., for finding electrostatic potential in different situations are bombarded so quickly on students that they do not have sufficient opportunity to organize their knowledge. The given problem can be effectively solved using Laplace's equation as there are no charges in the channel, where the potential is to be calculated. However, while sensemaking, G4 introduced image point charges in the absence of any point charges given in the problem in the region of interest. They were thoughtful and stated that their aim was to place these image charges such that the boundary conditions are satisfied on both the upper and lower grounded conducting planes. G4 also reasoned that the channel between the grounded conducting planes consists of horizontal equipotential surfaces which is a very productive line of reasoning based upon their earlier inferences about the different potentials on the two grounded conducting planes. These instances provide valuable insight into how grad students who are developing expertise in these advanced concepts may activate knowledge resources in a context (in which they may or may not be relevant) while solving a problem while ensuring local coherence. G4's sensemaking shows local coherence throughout although it is influenced by what they contemplated at an earlier stage of sensemaking and lacks global coherent.

### C. Case C. Curved electric field lines and appropriateness of LE method for finding potential

We now discuss the sensemaking of G4 for Q2 and Q3, which ask students to choose if solving LE is an effective method for finding the potential for the different problems posed. G4 focused on electric field lines to answer these questions related to the appropriateness of LE method. For different options of Q2 and Q3, G4 played the Pictorial Analysis and Physical Mechanism games in a nested manner, as per their perceived need for a diagrammatic representation or a physical explanation for clarifying the situation given. Their identified target concepts for the Pictorial Analysis game were the electric field lines for most of the configurations or the symmetry (for option III of Q3). For the development of the story in the Physical Mechanism game, they focused on analyzing the curving/bending of electric field lines in most of the situations given.

We begin with a discussion of G4's sensemaking on Q2. G4 started by reading the question carefully and then read option I for Q2 saying, '…[for] *which of the following situations is solving Laplace's equation an appropriate and effective method for finding the potential V?*' After that they started thinking aloud about when LE can be used. They were not sure about the conditions which would allow one to use LE in a given situation. G4 thought that Laplace's equation could be used to solve for the potential, e.g., in problems that had vacuum and said, '*…I guess…you would solve for the potential by using Laplace's equation but that's only…in a vacuum or places where gradient squared of a potential always equal to zero… So, I'm going to just say yes* [G4 said that the problem in Q2 option I can be solved effectively using LE] *for number I* [i.e., option I of Q2] *if I assume it's in a vacuum everywhere else*'. This example shows that when G4 was asked to think about which method would be effective for different problems in Q2 for finding the potential, they activated knowledge resources that involved several concepts but not necessarily those relevant to the problems posed.

After finishing their reasoning about option I of Q2, G4 started reading option II of Q2 and reflected on whether LE is an appropriate and effective method for finding the potential. They thought that considering the curving of the electric field lines would tell them if they could use LE or not. They said, '*A uniformly charged spherical volume, a spherical volume! That makes me think that we are going to use $\nabla^2 V$*'. After finishing reading the problem, they perceived a need to have a diagrammatic representation of the configuration to see clearly how field lines are curving. They decided to draw a diagram saying, '*A*

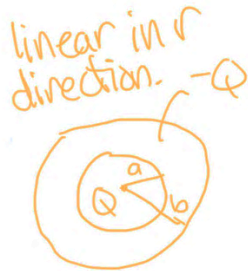

Fig. 6. Picture drawn by G4 for option II of Q2

*uniformly charged spherical volume. Okay, so this is not a conductor*'. They started drawing the diagram shown in Fig. 6, and said, '*radius 'a' and total charge 'Q'* [drew a sphere with radius '$a$' and wrote down charge '$Q$' as shown in Fig. 6], *enclosed with a concentric grounded, conducting spherical shell radius b > a and then this is '$-Q$'* [drew another concentric circle of radius '$b$' and labeled the charge as '$-Q$' as shown in Fig. 6]. They stated that the electric field lines do not bend in this case. Therefore, they started calling the problem in Fig. 6 a 'linear' problem because all electric field lines were radial and were not bending. They said, '*…that sounds like a pretty linear problem* [because electric field lines do not bend/curve]. *It feels linear according to radius, linear in the 'r'* [radial] *direction. So that to me doesn't feel like you would have to do some sort of partial derivative. So, I'm gonna say no to that one* [G4 said that the problem posed in option II of Q2 cannot be solved effectively using LE]'.

After this, G4 went back to option I of Q2 because they wanted to change their earlier response. They suddenly activated knowledge resources pertaining to the fact that they had solved a similar problem using the method of images before and therefore changed their earlier response for option I of Q2. They said, '*…And actually, I'm gonna go back, and I'm going to say 'no' to number I* [G4 said that option I in Q2 cannot be solved effectively using LE] *because I don't know why you would have to do a partial derivative if you're talking about a single charge here and then a conducting plate* [plane]*?*. They started drawing a charge and a grounded conducting plane as shown in Fig. 7 saying, '*…I feel like you can make the conducting plate* [plane] *disappear by using the method of images with two charges or something, one of them* [image charges] *could be '2q', and one of them could just be 'q' or '−q'…so that doesn't feel like it would be appropriate to use Laplace's equation*'. Thus, in this case [in option I of Q2], even though the electric field lines are curved, when G4 activated their knowledge resources that problems similar to the situation presented in option I can be solved using the method of images, they abandoned the idea that LE would be an appropriate and effective method in this situation since the field lines would be curved. This example also shows that the knowledge resources – for instance, the reasoning heuristic that G4 was using, in which the curving of the electric field lines implied that LE was appropriate for finding the potential – were abandoned (after being activated initially) after some other knowledge relevant to that problem took precedence. For example, in this case, G4 suddenly activated resources pertaining to the method of images being relevant for the situation in option I of Q2 (which is actually the most effective and efficient method for this situation), and even though they were somewhat unsure about what image charges would be relevant, they discarded their earlier thought that LE would be appropriate in this situation because the electric field lines curve.

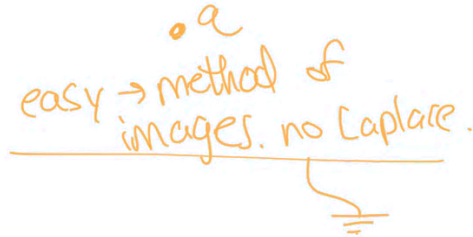

Fig. 7. Picture drawn by G4 for option I of Q2

Continuing with the problem, G4 read the problem posed in option III of Q2. After reading it, they started drawing the diagram saying, '*That sounds like a completely different problem than what we just did…so $y = 0$* [drew a line representing plane $y = 0$ as shown in Fig. 8] and *'a'* [drew a line representing plane $y = a$ as shown in Fig. 8] *and then they're connected at the plane $x = 0$* [drew a vertical line representing $x = 0$ plane] *but that plane now is held at $V_0$. There's a thin insulating layer at each corner, so the conductors do not touch…they are grounded. They're also conducting…*'. They started drawing the electric field lines to see if those lines curved. G4 said, '*…in this situation, if we're including fringe effects* [by 'fringe effects' they were

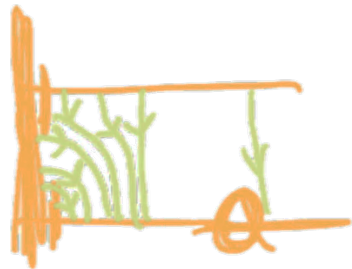

Fig. 8. Picture drawn by G4 to see the electric field lines for option III of Q2

referring to the curving of the field lines they drew in Fig. 8] *which I didn't really care about before, if we're really far away…the electric field lines would go from high to low* [potential] [drew vertical line pointing from the $y = a$ to the $y = 0$ plane, within the channel, away from the $x = 0$ plane in Fig. 8]. *And then nearby, it would…do some weird thing* [drew curved electric field lines near $x = 0$ plane as shown in Fig. 8]. *But they can't touch…but they always come at a perpendicular angle from conductors…And of course, it would depend on what the potential is at 'a'* [$y = a$] *but it would definitely curve. They are curving, so that sounds like a great opportunity to use* [LE], *E lines curve*'. Thus, they went back to their heuristic about LE being appropriate when field lines curve and chose that the problem given in option III of Q2 can be effectively solved using

LE. This choice was correct but their approach involving the curving of electric field lines was not correct. This points to the issue that using incorrect reasoning can sometimes lead to choosing the correct answer in multiple-choice questions.

It is noteworthy that this concept of the electric lines curving implying LE is appropriate for finding the potential was carried across multiple situations by student G4. We now discuss the sensemaking of G4 for another similar problem Q3. Q3 was asked in another session of the interview, but G4 continued using the same type of sensemaking involving the curving of electric field lines to decide whether LE is an effective method for finding the electric potential. G4 started by reading option I of Q3 carefully saying, '…*Okay, so…here's a spherical thing* [drew a spherical shell as shown in Fig. 9] , *and here's a point charge 'q'* [drew a point charge 'q' at a distance 'd'] *where this is at distance 'd', and this is 'R'* [radius of the spherical shell]. *So, it's a grounded spherical shell* [drew the grounding of the spherical shell]. *So, because it's grounded, charges can enter or exit that shell, so that charge on that shell is going to have a net charge equal and opposite to that point charge…I know that there would be more negative charges here* [drew negative induced charges on the spherical shell on the near side of the point charge 'q'] *and then more positive charges here* [drew positive induced charges on the spherical shell on the farther side of 'q']'. G4 did not realize that grounding will make positive charges go away and they thought that the spherical shell should be overall neutral because it is grounded. G4 then read the problem again and started thinking whether LE should be used here to find the potential saying, '…*I thought that we would use Laplace's equation, because it has a $\nabla^2$ in it, anytime we have $2d$ curvature of some sort in the E field lines* [we would use LE]. *So I guess in this situation if we're looking at the potential* [drew a point in space to look at the potential as shown in Fig. 9] *here then…E field lines would…occur in whatever way due to the charge on the sphere* [drew curved electric field lines due to induced charges on the sphere], *and then this charge here* [original charge 'q']*…it would be a summation of those* [electric field lines due to the original charge 'q' and the induced charges on the sphere]. *So, it's* [resultant electric field lines] *probably curved…so I'm going to say yes* [option I in problem Q3 can be effectively solved using LE to find the potential].

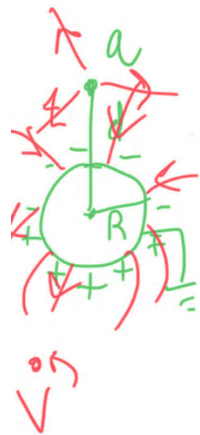

Fig. 9. Curved electric field lines drawn by G4 for option I of Q3

As noted, G4's sensemaking shows that they activated the concept of the curving of the electric field lines in different situations. In fact, as G4 proceeded through Q2 and Q3, they became reasonably confident that LE would be a good technique to find the potential when the field lines were curved. G4 then moved to the next part of the problem, read option II of Q3 saying, '…*there's something* [drew $y = 0$, $y = a$ and $x = 0$ planes as shown in Fig. 10]…*and they are very close to touching, but this is at some $V_0$* [labeled potential $V_0$ at $x = 0$ plane] *and then this is grounded, and this is grounded* [drew grounding for the $y = 0$ and $y = a$ planes]. *But just because it's grounded doesn't mean that it's a neutral charge, but because they're all conductors, it means that the change* [in potential] *across the surface is 0. But if we're here* [near the $x = 0$ plane], *it is* [potential] *going to be different than if we're here* [farther from the $x = 0$ plane]. *Far away…the E-field lines are going to be pretty parallel* [drew parallel electric field lines pointing from $y = a$ to $y = 0$ plane, farther from $x = 0$ plane]. *But nearby…they might curve because of that third* [vertical] *surface. So, I'm gonna say…we would probably need to use Laplace's equation there.*' G4 noted that they are able to draw the curved electric field lines in the diagram of the given problem at least for some regions [as shown in Fig. 10] so LE would be a good approach to find the potential in this situation.

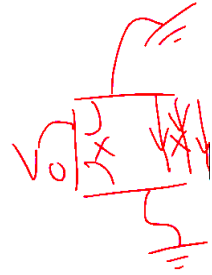

Fig. 10. Picture drawn by G4 for Q3 option II

After this, G4 read the next part, which is option III of Q3. They drew the diagram saying, '...*Okay, this one is a no* [i.e., option III of Q3 cannot be solved effectively using LE], *because if we have some thin disc* [drew a thin disc as shown in Fig. 11], *and we know the total charge* $\int \sigma \cdot dA = Q$, *and it's uniformly distributed then that sigma* [$\sigma$, charge density] *is a constant. And…if we are looking for the potential at any point* [drew a point in the space above the thin disc], *I feel like, instead of using* $V = \frac{kq}{r}$,…*we do* $\int \frac{k\sigma dA}{r}$ [$dA$ represents the area element on the disc]…*so I don't think that you would need to use Laplace's equation for that...because …it's very symmetric. So I feel like it* [potential] *can be done* [solved] *in spherical coordinates or cylindrical coordinates, I should say.*' At this point, G4 admitted that they were not very sure about when Laplace's equation should be used to find the potential saying, '*I just thought…that it was any time there's some 2d curvature in the E lines is when we're supposed to use Laplace's equation*'.

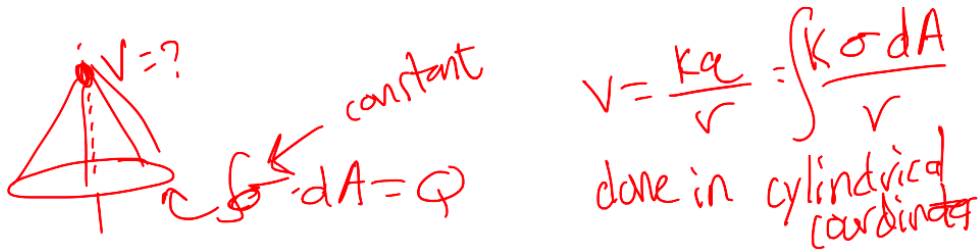

Fig. 11. Picture drawn by G4 for Q3 option III

In this case we observe that while solving Q2 and Q3, even though the problems ask whether LE is an effective method for finding electrostatic potential in different situations, G4 stepped back from considering the electrostatic potential to thinking about the electric field lines in the given situation. They used sensemaking involving the curving of the electric field lines to answer these questions because they had a vague idea that Laplace's equation problems somehow involve a 'curved (bent) electric field' or the curl of the electric field (they used these phrases curving/curling or the curvature of the electric field interchangeably at different points) and LE method is applicable if there are curved electric field lines involved. G4 also stated that if the electric field lines are 'linear' (by which they meant that they are not curved/bent), LE will not be applicable for finding the potential. It is not clear from G4's sensemaking why they thought this way and why they confused the curl of the electric field with curved electric field lines. We also note that the presence of curved electric field lines does not mean that LE is an appropriate method for finding the potential in the problem. Also, just because electric field lines bend in a given situation does not imply that the curl of the electric field is non-zero. In fact, in electrostatic situations, $\vec{E} = -\nabla V$ can be used to transform the differential form of Gauss's law into Poisson's equation (or Laplace's equation when the charge density in the region is zero). Also, the second derivative in each of $x$, $y$, or $z$ is related to the curvature. It appears that G4 somehow amalgamated these notions and came up with the heuristic that when the electric field lines curve/curl/have a curvature, LE is appropriate to find the potential.

G4's sensemaking while solving Q2 and Q3 regarding the options in which LE is appropriate for finding the potential provides valuable insight into this grad student's problem solving, their use of reasoning heuristics, and how an advanced physics student may weigh different knowledge resources that are activated while solving a problem and may abandon their reliance on reasoning heuristic they often use in cases in which other activated knowledge becomes more salient. For example, for Q2, Option I, in which a point charge was located above a grounded conducting plane, G4 initially invoked the incorrect heuristic that they used frequently in these types of problems that when the electric field lines curve, curl or have a curvature, LE is appropriate for finding the potential. However, after answering another question, they went back to Q2 Option I, saying they now recalled the fact that the method of images would be appropriate for this situation. This switching of responses shows how sensemaking evolves and as students activate different pieces of knowledge and reason with them, different resources may take precedence over others. For option II of Q2, G4 called it a 'linear' problem because the field lines were not curving (the lines were pointing radially from the sphere's center).

Also, G4 carried along the (incorrect) reasoning heuristic of curving of electric field lines implying that LE was appropriate for finding the potential from problem Q2 to Q3. G4's sensemaking shows the activation of different knowledge resources while solving different problems and different aspects of their sensemaking point to their evolving expertise. It is also praiseworthy that G4's sensemaking involved playing the Pictorial Analysis game for each option of both problems consistently. In particular, students were not explicitly asked to draw diagrams for any of the configurations but G4 engaged in sensemaking by choosing to visualize the configuration in the problem every time.

### D. Case D. Effective representation

Apart from the thoughtful sensemaking on Q1- Q3 discussed in previous cases, G4's sensemaking on problem Q4 was the most expert-like among all interviewed graduate students. Q4 is a challenging problem which asks about the electric field lines and electrostatic potential in a given situation. Below, we discuss how G4's response in this case was close to correct and better than all the other interviewed graduate students. Moreover, to shed light on G4's evolving expertise displayed on Q4, for comparison, we discuss the performance of other grad students.

Problem Q4 requires students to draw a diagram. This need to draw the diagram for the given configuration and show the electric field line and equipotential surfaces within the channel made G4 enter the Pictorial Analysis game [see Fig. 12]. Their identified target concepts were electric field lines and equipotential surfaces within the channel. After reading the problem statement, G4 said, '*Let me rewrite it for a second,* [wrote down $V = \sum_{n=1}^{\infty} \left(-\frac{4V_0}{n\pi}\right) e^{-\frac{n\pi x}{a}} \sin\left(\frac{n\pi y}{a}\right)$] *is the analytic solution for the potential expressed as an infinite series,* [is this mathematical expression] *helpful in making a sketch? I feel like that's a big…no, because instead of graphing a function, I'm trying to figure out what's happening to $n\pi$...But I guess if we were to go back, here's our y axis* [drew the $y$-axis as shown in Fig. 12] h*ere's our x-axis* [drew the $x$-axis as shown in Fig. 12]. *Here's $y = a$ right here, and this is the…side that is $V_0$* [drew a vertical line representing the $x = 0$ plane]…*We have blue* [line] *at top, this is grounded so $V = 0$* [drew the line representing plane $y = a$ and drew the grounding symbol]. *And then the red…one at the bottom was also held at $V = 0$* [drew the line representing plane $y = 0$ and drew grounding]'. G4 had evolved after solving different problems and did not appear to be getting confused with grounding anymore.

After drawing the planes, G4 focused on the rules to draw the electric field lines and equipotential surfaces saying, '*…Okay, so the rule that we have for the electric field lines is that electric field lines always…leave or enter a conductor at a right angle…But I would assume that all the electric field lines are gonna leave the green line* [plane $x = 0$] *and go towards the blue* [plane $y = a$] *and red* [plane $y = 0$] *because this* [$x = 0$ plane] *is some positive potential* [$V_0$] *and those* [$y = 0$ and $y = a$ planes] *are* [at] *zero* [potential]. *So if all of our electric field lines move from high to low…here's an electric field line* [drew an electric field line parallel to $y = 0$ plane at the center of the channel as shown in Fig. 12] *and I feel like here* [between the central electric field line and $y = a$ plane], *it would be more like this* [drew curved electric field lines as shown in Fig. 12]. *And then kind of mirror down here* [between the central electric field line and $y = 0$ plane] *kind of like that* [drew the curved electric field lines]…*because the electric field lines…point from high potential to low potential. So they point from $V_0$* [plane $x = 0$] *towards zero* [potential at surface $y = 0$ and $y = a$]. *And that also makes sense because at $x = \infty$, that's gonna be $V = 0$…I tried to draw an even concentration* [actually it should not be evenly concentrated] *of them* [electric field lines] *coming out of $V_0$* [$x = 0$ plane] *because across $V_0$, there's no change of potential…from one point to another. But the farther out in x we go, as $x \to \infty$, concentration of E field lines decreases. And so that means that potential goes towards zero. I think that makes sense.*'

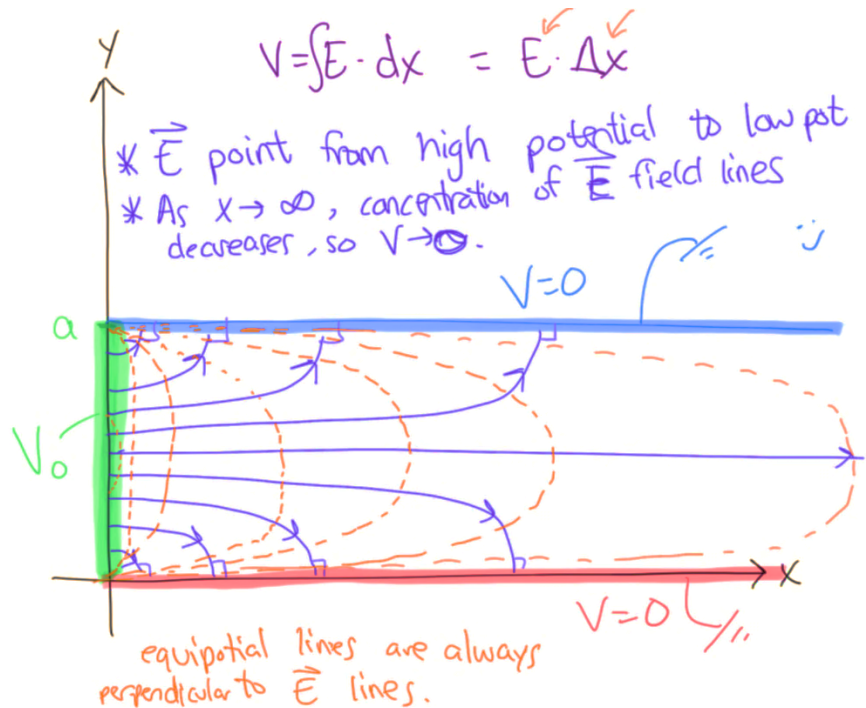

Fig. 12. A diagram drawn by G4 for Q4 which has many of the correct features

After this, G4 tried to draw the equipotential surfaces and said, '…*Okay, my equipotential surfaces, it's going to be…almost parallel to the y-axis* [drew first equipotential surface almost parallel to $x = 0$ near it] *when we're very close. The farther we go out, the more it's gonna be curved* [drew curved equipotential surfaces as shown in Fig. 12]. *Equipotential lines are always perpendicular to E field lines…Okay? So then it's almost like they get very flat out here. And also, none of these lines cross each other…there should be some sort of like increasing gap in here* [between the equipotential lines]'. Once G4 was satisfied with the reasoning of how the electric field lines and equipotential surfaces look like within the channel and were done drawing the detailed diagram as per expected in the question, they exited the Pictorial Analysis game.

In this case of G4's sensemaking about Q4, electric field lines should not be evenly concentrated as they come out of the plane $x = 0$, and should be highly concentrated near the corners. G4 did not take into account that field lines are not equidistant along the $y$ axis but they still drew the electric field lines and equipotential surfaces beautifully and almost correctly, which is highly praiseworthy considering this problem is very challenging.

To better appreciate G4's effective approach to this problem requiring diagrammatic representation, we now discuss how other grad students approached this problem Q4 and how different knowledge resources were activated for different students for the same problem Q4. As the problem demands a diagram, similar to G4, other grad students also played the Pictorial Analysis game.

For example, we now discuss the sensemaking of student G3, who approached Q4 very differently and focused on the analytic solution. The perceived need to draw the diagram of the given configuration and show the electric field lines and equipotential surfaces made G3 enter the Pictorial Analysis game. They wanted to use the concept of electric field and the analytical solution of the problem (they obtained) consisting of the sinusoidal function to understand how the potential and electric field lines vary within the channel. The final diagram drawn by G3 is shown in Fig. 13.

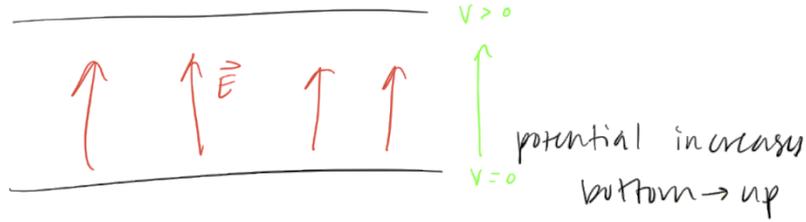

Fig. 13. Electric field lines drawn by G3 for Q4

After reading the question, G3 said, '…*I can't remember what equipotential lines looked like for planes; I remember for point charges. I'm gonna think about this in terms of the field lines because that'll tell me the values of the equipotential surfaces, if I know where the field lines are pointing towards. So here are my planes* [drew two parallel lines representing the $y = 0$ and $y = a$ planes in Fig. 13]. *Wait! How does this tell me about the potential…from increasing y, it has to be odd. So let me look at the* $\sin\left(\frac{n\pi y}{a}\right)$, $y > 0, n = odd$ [wrote it down], *y increases but n is also odd. How does sine behave?* [drew a sinusoidal function not shown]…*Okay, so if it's always gonna be plus or minus one if y is a. At $y = 0$, sine is 0…potential increases from the bottom plane to the top one…I don't think I know about the charges, but they're* [the conducting planes] *grounded. And I think if that's the case, then maybe I think, for a positive charge, the electric field line points in the direction of the increasing potential lines, equipotential lines. And then, if there's a negative charge, it would point away from it. So I think my field lines would also look like this* [drew the electric field lines pointing from bottom to top for the case where the top plane is at a higher potential in Fig. 13]. G3 thought about the behavior of sinusoidal functions but at the end only focused on the boundary conditions. G3 also incorrectly thought that the upper grounded conducting plane at $y = a$ has a higher potential value than the lower plane at $y = 0$ and therefore, drew the electric field lines pointing from the lower plane to the upper plane. In this example, we observed that G3 thought that the potential increases from top to bottom and the electric field lines point from the lower conducting plane to the upper conducting plane. They also did not show equipotential surfaces in the diagram that the problem asks for. Once they obtained the detailed figure of the given configuration, they exited the Pictorial Analysis game.

We now summarize the sensemaking of student G2 on Q4. Fig. 14 shows the diagram drawn by student G2 for equipotential surfaces G2 entered the Pictorial Analysis game immediately after reading the problem. G2 focused on the expression for the analytical solution to make the sketch. G2 said,

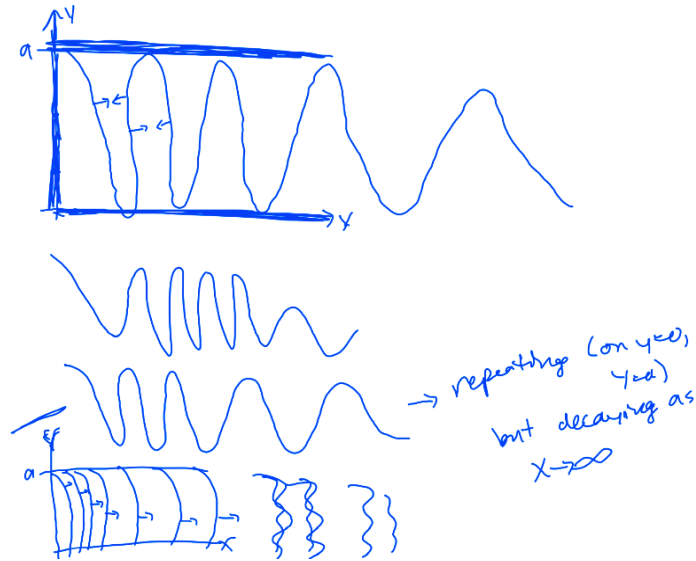

Fig. 14. Equipotential surfaces drawn by G2 for Q4

'*…I guess if we have our surface* [drew two conducting planes at $y = 0$ and $y = a$ along with the axes as shown in the first diagram of the Fig. 14]…*so repeating but decaying* [drew two oscillating functions under the top diagram, the second and third diagram in Fig. 14]...*I guess I'm gonna go with something* [like this] [drew the oscillating function within the channel in the first diagram in Fig. 14]…'. They decided to redraw their diagram [drew a fourth diagram in Fig. 14] saying that the first diagram in Fig. 14 might not be correct, '*maybe…they are close together at the beginning* [drew equipotential surfaces within the channel as shown in the fourth diagram in Fig. 14], *and then they get farther apart and maybe they have wiggles in between…and then get further apart, or maybe get lower amplitude, like, maybe we start with bigger and you continue…I guess that would make more sense…Okay, feels good, they're at least not…going back in on themselves. So maybe we'll go with this one…*'. In this case, we observe that G2 tried to use the sinusoidal functions they obtained in the expression (while solving for potential) to figure out how the equipotential surfaces should look within the channel. Their sensemaking shows the activation of valuable knowledge resources but they did not draw equipotential surfaces correctly as those should meet at the vertical plane $x = 0$, where the potential is $V_0$. G2 also did not draw the electric field lines. Once G2 obtained the detailed diagram for the given configuration, they exited the Pictorial Analysis game.

Another student G5 also approached problem Q4 by entering the Pictorial Analysis game. They used their intuition to draw the diagram and spent less time thinking about the details. They drew the channel and drew the equipotential surfaces as shown in the first diagram of Fig. 15 and said, '*...Actually, I think they're* [equipotential surfaces in the first diagram of Fig. 15] *supposed to get closer together, aren't they?...not sure they're supposed to be like how I originally drew them and the field lines. So this is the potential* [pointed to the first diagram in Fig. 15, representing the equipotential surfaces]...'. After that, they drew the electric field lines in the second drawing in Fig. 15 saying, '*…This is the E field I believe...and this is all intuition…I know that I want the potential to be constant over here* [pointing to $y = 0$ plane in the second diagram of Fig. 15], *and constant over on these ones too* [pointing to the $y = a$ plane in the second diagram of Fig. 15].' G5 added, '*Oh, well. I don't have enough intuition for field lines...and to be perfectly honest, these numerical solutions* [provided] *only make things worse*'. Thus, G5's drawings of the equipotential surfaces and electric field lines drawn using intuition are not close to the correct solution. Once G5 obtained the detailed diagrams they exited the Pictorial Analysis game.

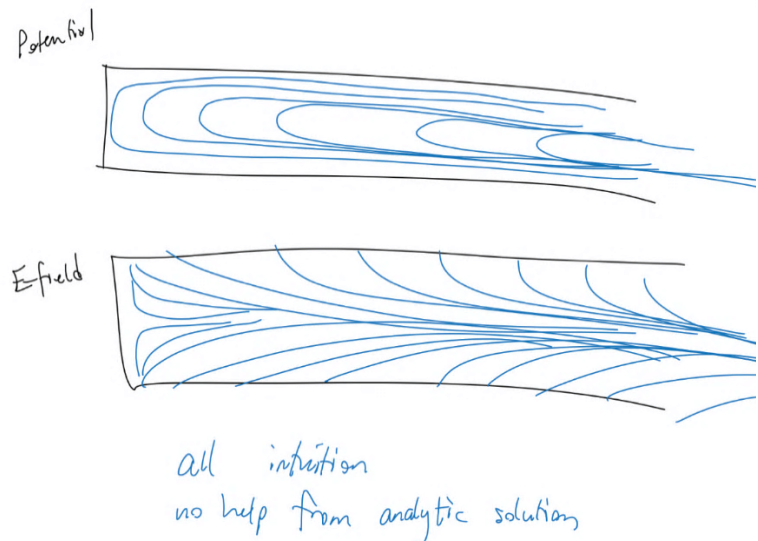

Fig. 15. Electric field lines drawn by G5 for Q4

In this case, we observed that Q4 was not an easy problem and students generally struggled with it but G4's solution was the most expert-like out of all the graduate students discussed. Although G4 used the concept of electric field lines incorrectly in Case C for problems Q2 and Q3 discussed earlier, the cognitive effort involved in contemplating about electric field lines (even though those problems only asked for the potential) appears to be productive for G4 in solving problem Q4 discussed in Case D. They were able to use their knowledge resources about the field lines very well for Q4 as they progressed through the tutorial. In particular, G4 was successfully able to draw the electric field lines and equipotential surfaces (correctly) for problem Q4 where other students generally struggled. This also shows that the activation of resources in the right context can be very constructive. G4 appeared to construct knowledge in the process of sensemaking and gained a better understanding of how to solve these problems as they progressed through them. It appears that unlike G2 and G3, G4 took advantage of the broader features of the solution as opposed to being excessively hung up on the analytic solution provided. Also, unlike G5 who completely relied on their gut feeling, G4's reasoning had a good blend of intuitive and analytic reasoning. Helping students blend their intuitive and analytic reasoning should be an explicit goal to help students become better problem solvers.

Advanced physics students' sensemaking here also points to the importance of developing representational fluency and focusing on helping students learn effective representations, e.g., by asking students to convert from one representation to another and asking them to analyze problems pictorially. Grad student sensemaking also underscores the importance of helping them develop the ability to integrate mathematical and physical concepts and reflect on the physical significance of different aspects of the complex problem posed.

## IV. SUMMARY

Prior research on student sensemaking while solving physics problems has primarily focused on introductory physics students. This research extends this important thread of research to physics graduate students' sensemaking primarily in the context of solving upper-level electrostatics problems with a focus on problems that can be solved using LE. This paper is primarily a case study of one grad student (G4) although we compare their sensemaking with other grad students' sensemaking in some cases to highlight G4's deep engagement and search for local coherence even if global coherence was not achieved. We mapped G4's sensemaking onto different epistemic games of Tuminaro and Redish [63]. Although the sensemaking of G4 often did not show global coherence, across all problems, their consistent quest for local coherence during sensemaking is insightful and praiseworthy.

We found that different epistemic games were frequently nested. To our knowledge, nesting has not been discussed within the framework of epistemic games in introductory physics problem-solving literature. However, nesting of games is likely to be common in physics problem-solving at both introductory and advanced levels depending on the complexity of the problems and whether one analyses sensemaking for a long segment of the problem-solving process or the entire process as in our case. The detailed sensemaking episodes we discussed here highlight the feature of nesting. In particular, we observed that while sensemaking, students switched between different games or started playing a new game within the moves of another game. Thus, the epistemic games [63] they played often had a nested structure. Furthermore, depending upon researchers' perspective, one of the moves, e.g., of the mapping meaning to mathematics game (e.g., developing/evaluating a story) itself can consist of a physical mechanism game. Overall, G4's sensemaking shows local coherence throughout even if global coherence was challenging.

In Case A, which focuses on the 'Conflict about values of potential on two parallel grounded conducting planes', the linear potential profile given on the vertical surface at $x = 0$ in Q1 confused G4 and made them conclude that there can be different constant values of potential on the upper and lower grounded conducting planes. G4's thoughtful sensemaking to resolve the conflict shows that they chose these potential values via careful reflection on various features of the problem and tried to validate their interpretation by activating different knowledge resources. They also concluded that the surface at $x = 0$ could not be a conductor and there should be some charge distribution on the surface even though the problem did not ask them about these issues. Some highlights of G4's sensemaking for case A are shown in Table 2.

Table 2. Highlights of G4's sensemaking for case A

| Case A: Perceived contradiction in what should be the potential on the upper grounded conducting plane in Q1 |
| --- |
| <ul><li>Since the upper conducting plane is grounded, it could be $V = 0$ (same as the lower grounded plane) but since it meets the surface at $x = 0$ which has a potential profile $V = \frac{V_0 y}{a}$, it could be $V_0$.</li><li>G4 focused on physical mechanism<ul><li>♦ Why the surface at $x = 0$ cannot be a conductor<ul><li>➢ Conductors are equipotential.</li><li>➢ Potential varies on the surface at the $x = 0$ plane as $V = \frac{V_0 y}{a}$.</li></ul></li></ul></li></ul> |

> ➤ This varying potential on the surface at $x = 0$ plane may be due to charge distribution on a non-conductor.
> ♦ Potential on the upper grounded conducting plane is $V = V_0$
>   ➤ Potential cannot have discontinuity.
>   ➤ Potential on different grounded conducting planes can be different.

In Case B, which focuses on 'Blending Laplace's equation with the method of images', we discussed the continuation of Case A. The focus on grounded conducting planes may have prompted G4 to think about the method of images (MoI) as the approach to solving problem Q1, even though it is not appropriate. They tried to introduce image charges such that the boundary conditions on the two parallel grounded planes $y = 0$ and $y = a$ are satisfied. Although this is not the correct method, they spent a significant amount of time finding image charges which (they thought) could be used to write the expression for potential within the channel. Case B shows that G4 mixed up different methods to solve problems in different contexts. A possible reason for such a mix up between different methods is that students learn about several approaches to finding electrostatic potential in a short interval in advanced physics courses. In Case B, G4 used the method of images for the problem which can be solved effectively using LE. This type of mix up is analogous to the case of another graduate student [94], who in another study used LE when the potential can be found effectively using the method of images. When students learn different concepts in short intervals of time, it can lead them to mix different concepts. Therefore, even advanced physics students should be given sufficient time to learn one approach before moving to another (and also be provided scaffolding support in contrasting their applicability conditions). G4 also drew analogy between the problem posed and a parallel plate capacitor and thought that the potential within the channel varies linearly as $V = \frac{V_0 y}{a}$ as there are equipotential surfaces within the channel, along the surfaces parallel to the conducting planes. This analogy with the parallel plate capacitor is an excellent example of their thoughtful sensemaking even though it is only appropriate for the boundary conditions they convinced themselves of (the upper grounded conducting plane is at potential $V = V_0$ and the lower grounded conducting plane is at potential zero). The highlights of G4's sensemaking for Case B are shown in Table 3.

The comparison of G4's sensemaking with G3's sensemaking discussed in Case A for Q1 also shows how these two grad students, who appeared to be proposing similar solutions, had a significantly different depth in their sensemaking. G4 and G3 both interpreted that there should be two different potential values for the two parallel grounded conducting planes due to confusion with the potential profile on the surface at $x = 0$. However, the two evolved differently in the sensemaking process after this step. Only G4 did deep sensemaking to resolve the conflict (e.g., whether the potential on the upper grounded conducting plane is $V_0$ or zero) by examining different aspects of the problem carefully to understand the situation. G3's approach seems to be somewhat simplistic, and without worrying about whether the potential on the upper grounded conducting plane can be $V_0$ (which is different from the lower grounded conducting plane at zero potential), they quickly concluded that was the case. Also, G3 immediately concluded that the potential should be $V = \frac{V_0 y}{a}$ within the channel (this expression does not have an $x$ dependence) without reflecting on the fact that their proposed solution was too simple ($V = \frac{V_0 y}{a}$ at $x=0$ was provided in the problem statement). In particular, G3 did not engage in deep sensemaking by questioning how the solution of the problem could be so trivial considering the potential profile $V = \frac{V_0 y}{a}$ at $x = 0$ was provided in the problem statement. On the other hand, G4 demonstrated deep sensemaking at every step and they tried to justify each move. Although G4 also eventually concluded that the potential within the channel should be $V = \frac{V_0 y}{a}$, their rich sensemaking involved reflection at each step (including a great analogy between the channel and a parallel plate capacitor with plates at different potentials) and showed local coherence throughout in their reasoning. G4's and G3's conclusion that the two grounded conducting planes are at different potentials also sheds light on the extent to which advanced students understand what grounding a conductor means without any prompting.

Table 3. Highlights of G4's sensemaking for Case B

> Case B: Perception that the method of images is an effective approach to find the potential in the channel and image charges must be placed to satisfy the boundary conditions in Q1

- We hypothesize that G4 decided that method of images is an effective approach because grounded conducting planes are often salient in method of images. They had spent a significant amount of time on figuring out the potential on the upper grounded conducting plane.
- G4 placed four image charges (although no point charges are present in the problem statement) that they thought will help satisfy the boundary conditions at the upper and lower grounded conducting planes.
  - These charges were placed to ensure that the potential was zero at the lower grounded conducting plane.
  - They wanted to ensure that the total potential on the upper grounded conducting plane was $V_0$ using the sum of the potential due to four image point charges and their distances from the plane to satisfy the boundary conditions they perceived to be correct.
- G4 drew an excellent analogy between their diagrammatic representation of the situation in problem Q1 (the upper grounded conducting plane at potential $V = V_0$ and the lower grounded conducting plane at potential zero) and a parallel plate capacitor.
  - Concluded that the channel where they were figuring the potential consists of equipotential surfaces parallel to the upper and lower conducting planes consistent with $V = \frac{V_0 y}{a}$.
  - Used the value of constant $V_0$ that they found using image charges in the expression $V = \frac{V_0 y}{a}$.
- G4 assumed that the problem was solved at this point.
- G4 strived for local coherence throughout while navigating different components of the problem-solving process (comparison shows that G3 did not do deep sensemaking even though they arrived at the same solution; G3 did not reflect upon their problem solution. G3's solution is trivial and does not require deep analysis in that $V = \frac{V_0 y}{a}$ provided for $x = 0$ in the problem statement was their potential everywhere in the channel).

In Case C, which focuses on 'Curved electric field lines and appropriateness of LE method for finding potential', G4 used a reasoning heuristic that LE can be effectively used to solve a problem if electric field lines curve (or curl or had curvature, words they sometimes used interchangeably) for a given charge configuration. Although we are not certain why they used this heuristic, it is possible that they got confused with how in the electrostatic situation, the fact that the curl of the electric field is zero can be used (along with the electric field being the negative of the gradient of the potential) to turn the differential form of Gauss's law into the Poisson's equation for the potential (or LE if there is no charge in the region). G4 thought that when the electric field lines curve, LE can be used to find the potential in various parts of problems Q2 and Q3. What is also noteworthy is that they used this reasoning heuristic to decide if LE is an effective method to find the potential or not but if other knowledge resources were activated in different subproblems of Q2 and Q3 (e.g., resources related to the fact that the method of images or direct integration method may be a better approach to finding the potential), they gave precedence to the approach that seemed most appropriate in a particular context. We also find that G4 sometimes seemed to be answering a question correctly about whether LE would be an appropriate approach to finding the potential even though the reasoning was incorrect. This situation can occur for multiple-choice tests and shows that it can be difficult for the instructor to recognize whether a student is choosing the correct answer using the correct reasoning. The highlights of G4's sensemaking for case C are shown in Table 4.

Table 4. Highlights of G4's sensemaking for case C

| Case C: Conversion from verbal to diagrammatic representation with detailed diagrams including field lines to figure out if LE is an effective approach to find potential in Q2 and Q3 |
| --- |
| - G4 used a heuristic about field lines curving, curling or having curvature being related to LE being an effective approach to find potential.<br>  - Stepped back from the potential (that the problem asked for) to sensemaking about field lines.<br>  - Heuristics about curved field lines were used for all problems unless knowledge resources were activated pertaining to another more effective approach in a given context. |
| ♦ In some cases, G4 went back to revising their solution, which shows reflection. |

In case D, which focuses on 'Effective representation', G4 used their knowledge resources related to drawing electric field lines in problem Q4 beautifully and almost correctly drew electric field lines and equipotential surfaces for the given problem. We compared G4's approach with other grad students' approaches to Q4 to show that this was a very challenging problem and G4's diagrammatic representation was the most accurate of all. The highlights of G4's sensemaking for Case D are shown in Table 5. Some of the other students we compared G4's sensemaking with activated useful resources and had reasonable representations but struggled at different stages. Even though different grad students had diagrammatic representations of electric field lines and potential that looked very different from each other, they each activated valuable knowledge resources for Q4. However, other grad students typically either used their gut feeling or the complex analytic expression (but not both that G4 beautifully combined) to guide their drawings.

Table 5. Highlights of G4's sensemaking for Case D

| |
|---|
| Case D: G4 used a combination of intuitive knowledge as well as formal prior knowledge and new knowledge generated during sensemaking for nearly correct diagrammatic representations of electric field lines and potential in Q4. |
| • G4's approach was nearly correct for a very challenging problem in which other grad students struggled and either only used their gut feeling or the complex analytic expression (which wasn't necessarily helpful). |

In this case study, advanced students' sensemaking shows their evolving expertise similar to introductory physics students' sensemaking in problem-solving appropriate for each level. Their sensemaking was often only locally coherent and lacked the data that support the findings of this article are not publicly available them. Graduate student G4 used reasoning heuristics, e.g., related to electric field lines curving implying LE can be used to find the potential similar to reasoning heuristics observed by Tuminaro and Redish in introductory physics context [63]. Moreover, in many cases, G4 dug deep into the problem going beyond what was explicitly asked in the problems (e.g., figuring out that the surface at $x = 0$ in problem Q1 cannot be a conductor under electrostatic conditions) to make sense of the problems. All the cases discussed are valuable to understand advanced student sensemaking so that we can develop curricula and pedagogies and provide them with appropriate scaffolding support to help them develop their expertise further.

### ACKNOWLEDGMENTS

We acknowledge the foundational contributions of Joe Redish, whose work has profoundly influenced physics education research and the framing of this study. We thank the students who participated in this study. We thank Dr. Robert P. Devaty for his constructive feedback on the manuscript.

### DATA AVAILIBILITY

The data that support the findings of this article are not publicly available because they contain sensitive personal information. The data are available from the authors upon reasonable request.

---